\documentstyle[11pt,epsf]{article}
\newcommand{\postscript}[2]
   {\setlength{\epsfxsize}{#2\hsize}
   \centerline{\epsfbox{#1}}}
\newtheorem{theorem}{Theorem}

\newtheorem{assumption}{Assumption}

\newcommand{\x}{{\mathbf x}}

\newcommand{\bi}{\begin{itemize}}
\newcommand{\ei}{\end{itemize}}
\newcommand{\be}{\begin{equation}}
\newcommand{\ee}{\end{equation}}
\newcommand{\bea}{\begin{eqnarray}}
\newcommand{\eea}{\end{eqnarray}}
\newcommand{\ba}{\begin{array}}
\newcommand{\ea}{\end{array}}
\newcommand{\D}{\Delta}

\begin{document}
\def\theequation {\arabic{equation}}
\makeatletter\@addtoreset {equation}{section}\makeatother

\begin{center}
{\bf\Large Universally-convergent Squared-operator Iteration Methods for Solitary
Waves in General Nonlinear Wave Equations}

Jianke Yang$^{\dagger, \dagger\dagger}$ and Taras I.
Lakoba$^\dagger$

{$\dagger$: \small Department of Mathematics and Statistics,
University of Vermont, Burlington, VT 05401  \\
$\dagger\dagger$: \small Zhou Pei-Yuan Center for Applied
Mathematics, Tsinghua University, Beijing, China}
\end{center}




Summary:

Three new iteration methods, namely the squared-operator method, the
modified squared-operator method, and the power-conserving
squared-operator method, for solitary waves in general scalar and
vector nonlinear wave equations are proposed. These methods are
based on iterating new differential equations whose linearization
operators are squares of those for the original equations, together
with acceleration techniques. The first two methods keep the
propagation constants fixed, while the third method keeps the powers
(or other arbitrary functionals) of the solution fixed. It is proved
that all these methods are guaranteed to converge to any solitary
wave (either ground state or not) as long as the initial condition
is sufficiently close to the corresponding exact solution, and the
time step in the iteration schemes is below a certain threshold
value. Furthermore, these schemes are fast-converging, highly
accurate, and easy to implement. If the solitary wave exists only at
isolated propagation constant values, the corresponding
squared-operator methods are developed as well. These methods are
applied to various solitary wave problems of physical interest, such
as higher-gap vortex solitons in the two-dimensional nonlinear
Schr\"odinger equations with periodic potentials, and isolated
solitons in Ginzburg-Landau equations, and some new types of
solitary wave solutions are obtained. It is also demonstrated that
the modified squared-operator method delivers the best performance
among the methods proposed in this article.

\section{Introduction}

Solitary waves play an important role in nonlinear wave equations.
While such waves in some wave equations can be obtained analytically
(such as in integrable equations), they defy analytical expressions
in most other cases. Thus numerical computations of solitary waves
is an important issue. In the past, a number of numerical methods
have been developed for solitary waves. One of them is Newton's
iteration method. This method can converge very fast. However, it
often employs the finite-difference discretization, which has a low
accuracy (compared to spectral or pseudospectral methods). In
addition, in higher dimensions, time-efficient programming of this
method becomes more cumbersome. Recent studies also showed that this
method can suffer erratic failures due to small denominators
\cite{Boyd}. Another method is the shooting method (see
\cite{YangPRE02} for instance). This method works for all
one-dimensional problems and higher-dimensional problems which can
be reduced to the one-dimensional problem (by symmetry reduction,
for instance). It is efficient and highly accurate. However, it
fails completely for higher-dimensional problems which are not
reducible to the one-dimensional problem. A third method is the
nonlinear Rayleigh-Ritz iteration method (also called the
self-consistency method) \cite{Segev_consistency, Panos}, where one
treats the nonlinear eigenvalue problem as a linear one with a
solution-dependent potential. The solitary wave is obtained by
repeatedly solving the linear eigenvalue problem and normalizing the
solution. This method can become cumbersome as well in high
dimensions when the linear eigenvalue problem becomes harder to
solve. Two more methods are the Petviashvili method \cite{Pet,
Peli_Pet, MussYang, AbMuss} and the imaginary-time evolution method
\cite{GarciaRipollP_01, Bao_Du, Bao_04, VS_04, YangLakoba}. The
convergence properties of the former method were studied in
\cite{Peli_Pet} for a class of equations with power nonlinearity,
while those of the latter method were obtained in \cite{YangLakoba}
for a much larger class of equations with arbitrary nonlinearity.
Interestingly, it was shown that the convergence of the latter
method is directly linked to the linear stability of the solitary
wave if the solitary wave is sign-definite \cite{YangLakoba}. Both
methods converge fast, are highly accurate, are insensitive to the
number of dimensions, and their performances are comparable
\cite{YangLakoba}. However, both methods diverge for many solitary
waves (especially ones which cross zeros, i.e., excited states)
\cite{Peli_Pet, YangLakoba}. In recent years, some interesting yet
challenging solitary wave problems arise in physical applications.
Two notable examples are nonlinear light propagation in
multi-dimensional periodic and quasi-periodic media, and
Bose-Einstein condensates loaded in multi-dimensional harmonic
magnetic traps and periodic optical lattices. These problems are not
reducible to one-dimensional problems, so the shooting method can
not be used. In addition, solitary waves in these problems often
cross zeros (as is always the case in higher bandgaps), thus the
Petviashvili method and the imaginary-time evolution method do not
converge. Furthermore, the numerical stability analysis of such
solutions require the solutions themselves to have high accuracy,
which is often hard to achieve by the Newton's method or the
self-consistency method. Thus new numerical schemes are called upon.
These schemes should be time-efficient, highly accurate, insensitive
to the number of dimensions, and capable of computing all types of
solitary waves in any scalar or vector nonlinear wave equations.
Equally importantly, these schemes should be easy to implement. None
of the previous methods meets all these requirements.

In this paper, we propose several classes of iteration methods for
solitary waves which do meet all the requirements described above.
These methods are based on two key ideas. The first key idea is to
iterate a modified equation whose linearization operator is a square
of that for the original equation. This idea (in a different form)
was first presented in \cite{GarciaRipollP_01}. Specifically, those
authors proposed to obtain excited-state solitary waves as
stationary points of a certain functional that is different from the
usual Lagrangian. In the present study, we will show that this
procedure is equivalent to using the aforementioned squared operator
in the iteration equation. We further show that this operator
structure guarantees that the proposed methods converge for all
types of solitary waves in any nonlinear wave equations. These
iteration methods are compatible with the pseudo-spectral method for
the computation of spatial derivatives, thus they are highly
accurate (with errors that decrease exponentially with the decrease
of the spatial grid size), and can handle multi-dimensional problems
with little change in the programming. The second key idea is to
introduce an acceleration operator to the scheme
\cite{GarciaRipollP_01, YangLakoba}. The acceleration operator
speeds up the convergence of these iteration methods drastically,
hence making them highly time-efficient. Based on these two key
ideas, we propose two powerful new iteration methods which we call
the squared-operator method and the power(or arbitrary
quantity)-conserving squared-operator method. The former method
specifies the propagation constant of the solitary wave, while the
latter method specifies the power or other arbitrary functional of
the solution. Both methods are shown to converge to any solitary
wave if the time-step parameter in the methods is below a certain
threshold value, and the initial condition is sufficiently close to
the exact solution. Beyond these two ideas, we also employ an
eigenmode-elimination technique \cite{LY06} which speeds up the
convergence of iterations even further. The resulting numerical
method will be called the modified squared-operator method in the
text. All these schemes pertain to solitary waves which exist at
continuous propagation constants. In certain nonlinear wave problems
(especially of dissipative nature), however, solitary waves exist at
isolated propagation constants. By extending the above ideas, we
construct the squared-operator method and the modified
squared-operator method for isolated solitary waves as well. By
applying these new schemes to various challenging solitary wave
problems, we demonstrate that the modified squared-operator method
gives the best performance, followed by the squared-operator method,
then followed by the power-conserving squared-operator method. In
the end, we construct a whole family of iteration schemes which
share similar convergence properties and contain the presented
schemes as special cases.

This paper is structured as follows. In Section 2, we present the
squared-operator method and show that it converges for all types of
solitary waves in arbitrary vector equations. We also demonstrate it
on the nonlinear Schr\"odinger equation where we carry out the
explicit convergence analysis and determine optimal parameters in
the acceleration operator. In Section 3, we present the modified
squared-operator method and show that this method not only converges
for all types of solitary waves in arbitrary equations, but also
converge faster than the squared-operator method. In Section 4, we
present power-conserving versions of the squared-operator method and
prove their universal convergence properties. In Section 5, we
derive the squared-operator method and the modified squared-operator
method for isolated solitary waves. In Section 6, we demonstrate all
these methods on various challenging examples, including
vortex-array solitons in higher bandgaps and solitary waves in
coherent three-wave interactions, and show that the modified
squared-operator method delivers the best performance. In the
Appendix, we present whole families of squared-operator-like
methods, and show that the methods described in the main text are
the leading members in these families in terms of convergence
speeds.

\section{The squared-operator method for solitary waves in general nonlinear wave equations}
\label{sec_SO}

Consider solitary waves in a general real-valued coupled
nonlinear wave system in arbitrary spatial dimensions, which can be
written in the following form
\begin{equation} \label{L0U}
\textbf{L}_0 \textbf{u}(\x)=0.
\end{equation}
Here $\textbf{x}$ is a vector spatial variable,
$\textbf{u}(\textbf{x})$ is a real-valued vector solitary wave
solution admitted by Eq. (\ref{L0U}), and $\textbf{u} \to 0$ as
$|\textbf{x}| \to \infty$. Note that for complex-valued solitary
waves, the equation can be rewritten in the above form with
$\textbf{u}$ containing the real and imaginary parts of the complex
solution. Let $\textbf{L}_1$ denote the linearized operator of Eq.
(\ref{L0U}) around the solution $\textbf{u}$, i.e.,
\begin{equation} \label{L1}
\textbf{L}_0
\left(\textbf{u}+\tilde{\textbf{u}}\right)=\textbf{L}_1\tilde{\textbf{u}}+O(\tilde{\textbf{u}}^2),
\hspace{0.5cm} \tilde{\textbf{u}} \ll 1,
\end{equation}
$(\cdot)^\dagger$ denote the Hermitian of the underlying quantity,
and $\textbf{M}$ be a real-valued positive-definite Hermitian
operator which we call the acceleration operator. The idea of our
method is to numerically integrate the time-dependent equation
\begin{equation} \label{ut}
\textbf{u}_t=-\textbf{M}^{-1}\textbf{L}_1^\dagger \textbf{M}^{-1}
\textbf{L}_0 \textbf{u}
\end{equation}
rather than $\textbf{u}_t=\pm \textbf{M}^{-1} \textbf{L}_0
\textbf{u}$. The reason is to guarantee the convergence of numerical
integrations, as we will demonstrate below. The operator
$\textbf{M}$ is introduced to speed up the convergence, in the same
spirit as the pre-conditioning technique in solving systems of
linear algebraic equations. Using the simplest time-stepping method
for Eq. (\ref{ut}) --- the forward Euler method, the iteration
method we propose for computing solitary waves $\textbf{u}$ is
\begin{equation} \label{SOM}
\textbf{u}_{n+1}=\textbf{u}_n-\left[\textbf{M}^{-1}\textbf{L}_1^\dagger
\textbf{M}^{-1} \textbf{L}_0
\textbf{u}\right]_{\textbf{u}=\textbf{u}_n}\Delta t.
\end{equation}
It will be shown below that this method is universally convergent as
long as the time step $\Delta t$ is below a certain threshold, and
this universal convergence stems from the fact that the iteration
operator for the error function is
$-\textbf{M}^{-1}\textbf{L}_1^\dagger\textbf{M}^{-1} \textbf{L}_1$,
or "square" of the operator $\textbf{M}^{-1} \textbf{L}_1$. Thus we
call scheme (\ref{SOM}) the squared-operator method (SOM). If
$\textbf{M}$ is taken as the identity operator (no acceleration),
then SOM (\ref{SOM}) reduces to
\begin{equation} \label{SOM0}
\textbf{u}_{n+1}=\textbf{u}_n-\left[\textbf{L}_1^\dagger
\textbf{L}_0 \textbf{u}\right]_{\textbf{u}=\textbf{u}_n}\Delta t,
\end{equation}
which has a simpler appearance but converges very slowly.

It should be noted that even though more complicated time-stepping
methods (such as Runge-Kutta methods) can also be used to integrate
Eq. (\ref{ut}), they are actually less efficient than the forward
Euler method (\ref{SOM}), because the extra computations in them
outweighs the benefits they may have. Implicit methods are even less
attractive. The reason is that due to the acceleration operator
$\textbf{M}$, the time steps $\Delta t$ allowed by explicit methods
such as (\ref{SOM}) for numerical stability (or convergence) are not
small, thus the need for implicit methods vanishes.

Scheme (\ref{SOM}) is actually one of the many SOM's one can
construct using the same squared-operator idea. Indeed, in the
Appendix, we will present a whole family of SOM's, of which
(\ref{SOM}) is a particular member. We will show there that scheme
(\ref{SOM}) is the leading member in that family in terms of
convergence speeds.

Let us now remark on the relation of these methods to the functional minimization
method for Hamiltonian equations proposed in \cite{GarciaRipollP_01},
which has the following form:
\begin{equation}
\textbf{u}_t=-\frac{\delta}{\delta \textbf{u}} \int \left\|
\textbf{L}_0 \textbf{u} \right\|^2 d\textbf{x}.
\label{spanish_functional}
\end{equation}
Here $\delta / \delta \textbf{u}$ represents the functional
(Frechet) derivative, and $\|...\|$ denotes the $L_2$-norm. Upon
taking this functional derivative and noticing that for Hamiltonian
equations, $ \textbf{L}_1^{\dagger} = \textbf{L}_1$, one recovers
Eq. (\ref{ut}) with $\textbf{M}=1$. The accelerated equation
(\ref{ut}) follows similarly from the functional equation
\begin{equation}
\hat{\textbf{u}}_t
 =-\frac{\delta}{\delta \hat{\textbf{u}}} \int \left\|
 \textbf{M}^{-1/2} \textbf{L}_0 \textbf{M}^{-1/2}
\hat{\textbf{u}} \right\|^2  d\textbf{x},
\label{accelerated_functional}
\end{equation}
where $\hat{\textbf{u}}=\textbf{M}^{1/2}\textbf{u}$.

To formulate the convergence theorem for the SOM (\ref{SOM}), we introduce the
following assumption.
\begin{assumption} \label{assumption0}
Let $\textbf{u}(\x; c_1, c_2, \dots, c_s)$ be the general solution
of Eq. (\ref{L0U}), where $(c_1, c_2, \dots, c_s)$ are free real
parameters. Then we assume that the only eigenfunctions in the
kernel of ${\bf L_1}$ are the $s$ invariance modes $\textbf{u}_{c_j}, 1\le
j \le s$, where $\textbf{u}_{c_j}=\partial \textbf{u}/\partial c_j$.
\end{assumption}

{\bf Remark 2.1} \ One example of these invariance modes is the translational-invariance mode.
Suppose Eq. (\ref{L0U}) is translationally invariant along the direction
 $x_1$, i.e., ${\bf u}(x_1+c_1, x_2, \dots, x_N)$ is a solution for any value of $c_1$.
Then the translational-invariance mode ${\bf u}_{c_1}={\bf u}_{x_1}$
is in the kernel of ${\bf L_1}$. Another example of the invariance
modes pertains to arbitrary phases of complex-valued solutions and
can be found, e.g., in Example 6.1 below.

{\bf Remark 2.2}\ Assumption 1 holds in the generic case. However,
in certain non-generic cases, it does break down (see Fig. 7 of
\cite{YangChen_defectsoliton} for an example).

Under Assumption 1, we have the following theorem.
\begin{theorem} \label{theorem0}
Let Assumption \ref{assumption0} be valid, and define
\be \label{dtmax} \Delta t_{max} \equiv -\frac{2}{\Lambda_{min}},
\ee
where $\Lambda_{min}$ is the minimum eigenvalue of operator
\begin{equation} \label{curlL}
{\cal L}\equiv
-\textbf{M}^{-1}\textbf{L}_1^\dagger\textbf{M}^{-1} \textbf{L}_1.
\end{equation}
Then SOM (\ref{SOM}) is guaranteed to converge to the solitary wave
$\textbf{u}(\x)$ of Eq. (\ref{L0U}) if $\Delta t < \Delta t_{max}$
and the initial condition is close to $\textbf{u}(\x)$. If $\Delta t
> \Delta t_{max}$, then SOM (\ref{SOM}) diverges from $\textbf{u}(\x)$.
\end{theorem}

\textbf{Proof.} We use the linearization technique to analyze SOM
(\ref{SOM}) and prove the theorem. Let
\begin{equation}  \label{linearize}
\textbf{u}_{n}=\textbf{u}+\tilde{\textbf{u}}_n, \qquad
\tilde{\textbf{u}}_n(\x)  \ll 1,
\end{equation}
where $\tilde{\textbf{u}}_n(\x)$ is the error. When Eq.
(\ref{linearize}) is substituted into SOM (\ref{SOM}) and only terms
of $O(\tilde{\textbf{u}}_n)$ are retained, we find that the error
satisfies the following linear equation:
\begin{equation} \label{error0}
\tilde{\textbf{u}}_{n+1}= \left( 1 + \Delta t \: {\cal L} \right)
\,\tilde{\textbf{u}}_{n},
\end{equation}
where ${\cal L}$ is defined in Eq. (\ref{curlL}).
Note that since ${\cal L}=\textbf{M}^{-1/2}{\cal
L}_h\textbf{M}^{1/2}$, where \[{\cal
L}_h=-\left(\textbf{M}^{-1/2}\textbf{L}_1\textbf{M}^{-1/2}\right)^\dagger
\left(\textbf{M}^{-1/2}\textbf{L}_1\textbf{M}^{-1/2}\right)\] is
Hermitian and semi-negative-definite, thus all eigenvalues of ${\cal
L}$ are real and non-positive. In addition, all eigenfunctions of
${\cal L}$ form a complete set in the square-integrable functional
space, hence $\tilde{\textbf{u}}_{n}$ can be expanded over ${\cal
L}$'s eigenfunctions. Consequently, if $\Delta t > \Delta t_{min}$,
the eigenmode with eigenvalue $\Lambda_{min}$ in the error will grow
exponentially due to $1+\Lambda_{min}\Delta t < -1$,  i.e., SOM
(\ref{SOM}) diverges. On the other hand, if $\Delta t < \Delta
t_{min}$, then no eigenmode in the error can grow. In fact, all
eigenmodes decay with iterations except those in the kernel of
${\cal L}$. But according to Assumption 1, eigenfunctions in the
kernel of ${\cal L}$ are all invariance modes $\textbf{u}_{c_j}$
which only lead to another solution with slightly shifted constants
$c_j$ and do not affect the convergence of iterations. Thus Theorem
\ref{theorem0} is proved.


\textbf{Remark 2.3}  \ The role of the acceleration operator
$\textbf{M}$ is to make $\Lambda_{min}$ of ${\cal L}$ bounded
(without $\textbf{M}$, $\Lambda_{min}=-\infty$ in general); see,
e.g.,  Example 2.1 below and \cite{YangLakoba}. As shown after
Remark 2.4 below, this leads to faster convergence of SOM
(\ref{SOM}).

\textbf{Remark 2.4} \ In computer implementations of SOM
(\ref{SOM}), spatial discretization is used. In such a case, if we
require the discretization $\textbf{M}^{(D)}$ of the
positive-definite Hermitian operator $\textbf{M}$ to remain
positive-definite and Hermitian (which is needed to show the
non-positiveness of eigenvalues of the discretized operator ${\cal
L}^{(D)}$), then Theorem \ref{theorem0} and its proof can be readily
extended to the spatially-discretized SOM scheme (\ref{SOM}), except
that $\Lambda_{min}$ in Eq. (\ref{dtmax}) becomes the smallest
eigenvalue of the discrete operator ${\cal L}^{(D)}$ now. Unlike
discretizations of some other schemes such as the Petviashvili
method \cite{Peli_Pet} and the accelerated imaginary-time evolution
method \cite{YangLakoba}, discretizations of SOM (\ref{SOM}) always
converge regardless of whether the discretized solitary wave is
site-centered (on-site) or inter-site-centered (off-site). Note that
under discretization, the kernel of ${\cal L}^{(D)}$ may become
smaller than that of ${\cal L}$ due to the breakdown of
translational invariances, but this does not affect the extension of
Theorem \ref{theorem0} and its proof to the discretized SOM scheme
at all.

In the application of SOM (\ref{SOM}), a practical question is how
to choose $\Delta t$ within the upper bound of $\Delta t_{max}$ so
that convergence is the fastest. To answer this question, let us
define the convergence factor $R$ of SOM (\ref{SOM}) as
\begin{equation} \label{R1}
R\equiv \max_{\Lambda}\left\{ \left| 1 + \Lambda \D t\right|
\right\},
\end{equation}
where $\Lambda$ is any non-zero eigenvalue of ${\cal L}$. Then the
error $\tilde{\textbf{u}}_n$ of the iterated solution decays as $R^n$, where $n$ is the
number of iterations. Smaller $R$ gives faster convergence. Clearly,
\be R \, = \, \max \left\{ \left| 1 + \Lambda_{min} \D t\right|, \,
 \left| 1 + \Lambda_{max}\D t\right| \right\},
\label{R2} \ee
where $\Lambda_{max}$ is the largest non-zero eigenvalue of operator
${\cal L}$. From this equation, we see that the smallest $R$
(fastest convergence) occurs at the time step
\be \D t= \D t_{*} \equiv -\frac2{\Lambda_{min}+\Lambda_{max}},
\label{t05_02} \ee
for which the corresponding convergence factor is
\be R_* = \frac{\Lambda_{min}-\Lambda_{max}}
{\Lambda_{min}+\Lambda_{max}}\,. \label{R*} \ee

Another practical question which arises in the implementation of SOM
(\ref{SOM}) is the choice of the acceleration operator $\textbf{M}$.
This operator should be chosen such that $\textbf{M}$ is easily
invertible. In addition, it should be chosen such that the
convergence is maximized or nearly maximized. A simple but often
effective choice is to take $\textbf{M}$ in the form of the linear
part of the original equation (\ref{L0U}), to which one adds a
constant to make it positive definite.

To demonstrate the effect of $\textbf{M}$ on convergence speeds,
below we consider the familiar nonlinear Schr\"odinger (NLS)
equation for which the convergence analysis of SOM can be done
explicitly.

{\bf Example 2.1} \ Consider the NLS equation in one spatial
dimension:
\begin{equation}\label{NLS}
u_{xx}+u^3=\mu u.
\end{equation}
Without loss of generality, we take $\mu=1$. For this $\mu$ value,
Eq. (\ref{NLS}) has a soliton solution
\begin{equation} \label{soliton}
u(x)=\sqrt{2}\: \mbox{sech}\: x.
\end{equation}
We take the acceleration operator $M$ to be in the form of the
linear part of Eq. (\ref{NLS}), i.e.,
\begin{equation} \label{M_form}
M=c-\partial_{xx},  \hspace{0.3cm} c > 0,
\end{equation}
which is easily invertible using the Fourier transform. Then the
eigenvalue equation for operator $M^{-1}L_1$,
\begin{equation}
\label{j3_24} M^{-1}L_1\psi=\lambda \psi,
\end{equation}
can be rewritten in the explicit form:
\begin{equation}  \label{NLSpsi}
\psi_{xx}-\frac{1+c\lambda}{1+\lambda}\psi+\frac{6}{1+\lambda}\mbox{sech}^2x
\psi=0.
\end{equation}
The continuous spectrum of this equation can be obtained by
requiring the eigenfunction to be oscillatory at $|x|=\infty$, and
this continuous spectrum is found to be
\begin{equation} \label{Lam_con} \ba{ll}
\lambda \in (-1, -\frac{1}{c}], & {\rm for}\;\; c>1; \vspace{0.5cm} \\
\lambda \in [-\frac{1}{c}, -1), & {\rm for} \;\; c<1. \ea
\end{equation}
The discrete eigenvalues of (\ref{NLSpsi}) satisfy the equation
\cite{Landau,Yang1996}
\begin{equation} \label{Lambdaeq}
\sqrt{\frac{1+c\lambda}{1+\lambda}}=\frac{1}{2}\left[\sqrt{1+\frac{24}{1+\lambda}}-(2j+1)\right],
\hspace{1cm} j=0, 1, 2, \dots,
\end{equation}
where the right hand side should be non-negative. These discrete
eigenvalues are plotted in Fig. 1(a). Note that $\lambda=0$ is
always a discrete eigenvalue with $j=1$. This zero eigenvalue is due
to the translational invariance of the NLS soliton (which leads to
$L_1u_x=0$). The eigenvalues $\lambda_0$ and $\lambda_2$ (for $j=0$
and 2) can be found to have the following expressions:
\begin{equation}
\lambda_0(c)=\frac{2(c+5)^2}{2c^2+9c+1+\sqrt{25c^2+118c+1}}-1,
\end{equation}
\begin{equation}
\lambda_2(c)=\frac{2(c+5)^2}{2c^2-3c+85+5\sqrt{25c^2-26c+145}}-1.
\end{equation}
From the above spectrum of operator $M^{-1}L_1$, we can easily
obtain the spectrum of the iteration operator ${\cal
L}=-M^{-1}L_1^\dagger M^{-1}L_1$, whose eigenvalues $\Lambda$ are
related to the eigenvalues $\lambda$ of $M^{-1}L_1$ as
$\Lambda=-\lambda^2$, since in this case $L_1^{\dagger}=L_1$. Hence
\begin{equation} \label{maxmin}
\Lambda_{max}(c)=-\lambda_2^2(c), \hspace{0.5cm}
\Lambda_{min}(c)=\mbox{min}\left\{-\lambda_0^2(c), -\frac{1}{c^2},
-1\right\}.
\end{equation}
These eigenvalues are plotted in Fig. 1(b). Based on these
eigenvalues, we can calculate the convergence factor $R_*(c)$ from
formula (\ref{R*}). This $R_*(c)$ is shown in Fig. 1(c). It is seen
that when $\lambda_0^2(c)=1$, i.e., $c=c_{opt}=6-\sqrt{13} \approx
2.4$, $R_*(c)$ reaches its minimum value $R_{min}\approx 0.80$. At
$c=c_{opt}$, $R_*(c)$ is not differentiable, because
$\Lambda_{min}(c)$ equals $-\lambda_0^2(c)$ for $c<c_{opt}$ and $-1$
(the edge of the continuous spectrum) for $c>c_{opt}$. For
$c=c_{opt}$, the dependence of the convergence factor $R$ on $\D t$
is shown in Fig. 1(d). At $\Delta t_*\approx 1.80$ from formula
(\ref{t05_02}), $R$ reaches its minimum value $R_{min}$ given above.
Without acceleration ($M=1$), this value of $R$ would be very close
to 1 with discretizations (since $\Lambda_{min}$ is large negative,
see \cite{YangLakoba}), and be exactly equal to 1 without
discretizations (since $\Lambda_{min}=-\infty$). From the above
explicit analysis, we see that the choice of $M$ affects the
convergence speed of SOM (\ref{SOM}) significantly.


\section{The modified squared-operator method for solitary waves in general nonlinear wave equations}

The above SOM (\ref{SOM}), even with a sensible choice of the
acceleration operator $\textbf{M}$, can still be quite slow for
certain problems (see the example in Fig. 4). In this section, we
employ an additional technique which can speed up the convergence of
SOM iterations even further. This technique is called eigenmode
elimination, and was originally proposed in \cite{LY06} for a
non-squared-operator scheme. When this eigenmode-elimination
technique is incorporated into SOM, the resulting method, which we
call as the modified squared-operator method (MSOM), will converge
faster than SOM (\ref{SOM}).

The MSOM we propose is
\begin{equation} \label{MSOM}
\textbf{u}_{n+1}=\textbf{u}_n-\left[
\textbf{M}^{-1}\textbf{L}_1^\dagger \textbf{M}^{-1} \textbf{L}_0
\textbf{u}- \alpha_n \langle \textbf{G}_n, \; \textbf{L}_1^\dagger \textbf{M}^{-1}
\textbf{L}_0 \textbf{u} \rangle \textbf{G}_n
\right]_{\textbf{u}=\textbf{u}_n}\Delta t,
\end{equation}
where
\begin{equation} \label{alpha}
\alpha_n= \frac{1}{\langle \textbf{M}\textbf{G}_n,\textbf{G}_n
\rangle}-\frac{1}{\langle \textbf{L}_1\textbf{G}_n,
\; \textbf{M}^{-1}\textbf{L}_1\textbf{G}_n\rangle\Delta t},
\end{equation}
$\textbf{G}_n$ is a function the user can specify, and the inner
product is the standard one in the square-integrable functional
space:
\begin{equation}
\langle \textbf{F}_1,
\textbf{F}_2\rangle=\int_{-\infty}^{\infty}\textbf{F}_1^\dagger
\cdot \textbf{F}_2 d\textbf{x}.
\end{equation}
If $\langle \textbf{F}_1, \textbf{F}_2\rangle=0$, $\textbf{F}_1$ and
$\textbf{F}_2$ are said to be orthogonal to each other. Two simple choices for
$\textbf{G}_n$ can be
\begin{equation} \label{choice1}
\textbf{G}_n=\textbf{u}_n,
\end{equation}
and
\begin{equation} \label{choice2}
\textbf{G}_n=\textbf{e}_n \equiv \textbf{u}_n-\textbf{u}_{n-1}.
\end{equation}

The motivation for MSOM (\ref{MSOM}) can be explained briefly as
follows \cite{LY06}. Consider SOM (\ref{SOM}), and denote the slowest-decaying
eigenmode of ${\cal L}$ in the error as $\textbf{G}(\textbf{x})$
with eigenvalue $\Lambda_s$.  Note that according to Eqs. (\ref{R1})
and (\ref{R2}), $\Lambda_s=\Lambda_{min}$ or $\Lambda_{max}$. Our
idea is to construct a modified linearized iteration operator for
the error so that eigenmode $\textbf{G}(\x)$ decays quickly, while
the decay rates of other eigenmodes of ${\cal L}$ remain the same.
If so, then this modified iteration scheme would converge faster
than the original SOM. For this purpose, consider the modified
linearized iteration operator
\begin{equation}   \label{calLmodified}
{\cal L}_M \Psi={\cal L}\Psi - \alpha \langle \textbf{MG}, \; {\cal L}\Psi
 \rangle \textbf{G},
\end{equation}
where $\alpha$ is a constant. Since $\textbf{G}(\x)$ is an
eigenfunction of ${\cal L}$, and recalling that eigenfunctions of
this ${\cal L}$ are orthogonal to each other under the
$\textbf{M}$-weighted inner product, we readily see that this
modified iteration operator and the original one have identical
eigenfunctions. Their eigenvalues are identical too except the one
for eigenmode $\textbf{G}(\x)$. The eigenvalue of this eigenmode
changes from $\Lambda_{s}$ of ${\cal L}$ to $(1-\alpha\langle
\textbf{MG}, \textbf{G} \rangle)\Lambda_{s}$ of ${\cal L}_M$. Then
if we choose $\alpha$ so that the decay factor of this eigenmode is
zero, i.e.,
\begin{equation}
1+(1-\alpha\langle \textbf{MG}, \textbf{G} \rangle)\Lambda_{s}\Delta
t=0,
\end{equation}
or equivalently,
\begin{equation} \label{alpha0}
\alpha= \frac{1}{\langle \textbf{MG},\textbf{G} \rangle}
\left(1+\frac{1}{\Lambda_{s}\Delta t}\right),
\end{equation}
then eigenmode $\textbf{G}(\x)$, which decays the slowest in SOM,
becomes decaying the fastest (in fact, this mode is eliminated from
the error after a single iteration), while decay rates of the other
eigenmodes in the error are unaffected. Thus convergence is
improved. In practical situations, the slowest-decaying mode
$\textbf{G}(\x)$ and its eigenvalue $\Lambda_s$ in SOM are not
known. In such cases, if we can choose $\textbf{G}_n$ which
resembles the slowest-decaying eigenmode of SOM, then the
corresponding eigenvalue $\Lambda_{s}$ can be approximated as the
Rayleigh quotient $\Lambda_{s}\approx \langle
\textbf{M}\textbf{G}_n, \; {\cal L}\textbf{G}_n \rangle / \langle
\textbf{MG}_n, \textbf{G}_n\rangle$. Substituting this approximation
into (\ref{alpha0}), we get $\alpha$ as given by formula
(\ref{alpha}). Corresponding to the modified linearization operator
(\ref{calLmodified}), the modified iteration method is then MSOM
(\ref{MSOM}).

The above derivation assumed that function $\textbf{G}_n$ is equal
or close to the slowest decaying eigenmode of SOM (\ref{SOM}). In
practical implementations of this method, one does not know priori
if the selected $\textbf{G}_n$ meets this criterion. This then may
put the effectiveness of this method in question. One may also ask
if this modified method can converge at all even with small time
steps. Fortunately, the convergence of MSOM is insensitive to the
choice of function $\textbf{G}_n$, at least when certain mild
conditions are satisfied. Specifically, let us expand function
$\textbf{G}_n (\textbf{u}_n)$ as
\begin{equation}
\textbf{G}_n=\textbf{G}_0+O(\tilde{\textbf{u}}_n),   \hspace{0.5cm}
\tilde{\textbf{u}}_n \ll 1,
\end{equation}
where $\textbf{G}_0$ is $\textbf{G}_n$'s leading-order term, and
$\tilde{\textbf{u}}_n$ is defined by Eq. (\ref{linearize}), then we
have the following theorem.

\begin{theorem} \label{theorem2b}
Let Assumption 1 be valid, and $\textbf{L}_1\textbf{G}_0 \equiv
\hspace{-0.35cm} / \hspace{0.2cm} 0$, then if
\begin{equation} \label{dtm}
\Delta t < \Delta t_M \equiv \min\left(-\frac{2}{\Lambda_{min}},
\frac{1}{\frac{\langle \textbf{M}\textbf{G}_0, \; {\cal L}\textbf{G}_0
\rangle}{\langle \textbf{MG}_0,
\textbf{G}_0\rangle}-\Lambda_{min}}\right),
\end{equation}
where ${\cal L}$ is given in Eq. (\ref{curlL}), $\textbf{M}$ is
Hermitian and positive definite, and $\Lambda_{min}$ is ${\cal L}$'s
minimum eigenvalue, then MSOM (\ref{MSOM}) is guaranteed to
converge to the solitary wave $\textbf{u}(\x)$ of Eq. (\ref{L0U}) if
the initial condition is close to $\textbf{u}(\x)$.
\end{theorem}

{\bf Proof.} \ We again use the linearization technique to analyze MSOM
and prove Theorem \ref{theorem2b}. Substituting (\ref{linearize})
into MSOM (\ref{MSOM}), we find that the error satisfies the linear
iteration equation
\begin{equation} \label{errorM}
\tilde{\textbf{u}}_{n+1}= \left( 1 + \Delta t \: {\cal L}_M \right)
\,\tilde{\textbf{u}}_{n},
\end{equation}
where
\begin{equation} \label{calLmPsi0}
{\cal L}_M\Psi ={\cal L}\Psi - \alpha_0 \langle \textbf{M}\textbf{G}_0, \; {\cal L}\Psi
\rangle \textbf{G}_0,
\end{equation}
and
\begin{equation}
\alpha_0=\frac{1}{\langle \textbf{MG}_0, \textbf{G}_0
\rangle}+\frac{1}{\langle \textbf{M}\textbf{G}_0, \; {\cal
L}\textbf{G}_0\rangle\Delta t}.
\end{equation}
Note that since $\textbf{M}$ is Hermitian and positive-definite, $\alpha_0$ is real.

Below, we show that all eigenvalues of ${\cal L}_M$ are real and
non-positive, the kernel of ${\cal L}_M$ is the same as that of
$\textbf{L}_1$, ${\cal L}_M$ has no square-integrable generalized
eigenfunctions at zero eigenvalue, and under the time-step
restriction (\ref{dtm}), $|1+\Lambda_M\Delta t|<1$ for all non-zero
eigenvalues $\Lambda_M$ of ${\cal L}_M$. Then under the assumptions
of Theorem \ref{theorem2b}, MSOM (\ref{MSOM}) will converge, and
Theorem \ref{theorem2b} will then be proved.

To proceed, we first write out the eigenvalue problem for operator
${\cal L}_M$, which is
\begin{equation} \label{calLmPsi}
{\cal L}\Psi - \alpha_0 \langle \textbf{M}\textbf{G}_0, \; {\cal
L}\Psi \rangle \textbf{G}_0=\Lambda_M \Psi.
\end{equation}
Taking the inner product between this equation and $\textbf{M}{\cal
L}\Psi$, and recalling the form (\ref{curlL}) of ${\cal L}$ and that
$\alpha_0$ is real and $\textbf{M}$ Hermitian, we see that
eigenvalues $\Lambda_M$ are all real. Next, we will analyze the
eigenvalue problem (\ref{calLmPsi}) by expanding $\Psi$ into
eigenfunctions of operator ${\cal L}$, a technique which has been
used before on other eigenvalue problems \cite{VK,Nachman}. First,
notice from the proof of Theorem 1 that eigenvalues of ${\cal L}$
are all real and non-positive, and eigenfunctions of ${\cal L}$ form
a complete set. Let the discrete and continuous eigenmodes of ${\cal
L}$ be
\begin{equation} \label{e-relation1}
{\cal L}\psi_k({\mathbf x})=\Lambda_k \psi_k({\mathbf x}),
\hspace{0.5cm} \Lambda_k \le 0,  \hspace{1cm} k=1, \dots, m,
\end{equation}
\begin{equation} \label{e-relation2}
{\cal L}\psi({\mathbf x}; \Lambda)=\Lambda \psi({\mathbf x};
\Lambda), \hspace{1cm} \Lambda \in I < 0,
\end{equation}
where $m$ is the number of ${\cal L}$'s discrete eigenvalues, $I$ is
${\cal L}$'s continuous spectrum, and the orthogonality conditions
among these eigenfunctions are
\begin{equation}
\langle \textbf{M}\psi_i, \psi_j \rangle = \delta_{i,j},
\label{ort_L1_d}
\end{equation}
and
\begin{equation}
\langle \textbf{M} \psi({\mathbf x}; \Lambda), \psi({\mathbf x};
\Lambda') \rangle = \delta(\Lambda-\Lambda'). \label{ort_L1_c}
\end{equation}
Then we expand $\textbf{G}_0$ and $\Psi({\mathbf x};\Lambda_M)$ as
\begin{equation} \label{u}
\textbf{G}_0=\sum_{k=1}^m c_k \psi_k({\mathbf x}) + \int_I
c(\Lambda)\psi({\mathbf x}; \Lambda)d\Lambda,
\end{equation}
\begin{equation} \label{psi}
\Psi({\mathbf x};\Lambda_M)=\sum_{k=1}^m b_k \psi_k({\mathbf x}) +
\int_I b(\Lambda)\psi({\mathbf x}; \Lambda)d\Lambda.
\end{equation}
Substituting Eqs. (\ref{u}) -- (\ref{psi}) into the eigenvalue
problem (\ref{calLmPsi}) and using the orthogonality conditions
(\ref{ort_L1_d}) -- (\ref{ort_L1_c}),  one obtains
\begin{equation} \label{cncs}
b_k=\frac{\alpha_0 c_k \langle \textbf{M}\textbf{G}_0, \; {\cal L}\Psi
\rangle}{\Lambda_k-\Lambda_M}, \hspace{1cm}
b(\Lambda)=\frac{\alpha_0 c(\Lambda) \langle \textbf{M}\textbf{G}_0, \; {\cal L}\Psi
\rangle }{\Lambda-\Lambda_M}.
\end{equation}
Notice that
\[
\langle \textbf{M}\textbf{G}_0, \; {\cal L}\Psi
\rangle=\sum_{k=1}^m\Lambda_k b_k c_k^* +\int_I
\Lambda b(\Lambda)c(\Lambda)^*d\Lambda
\]
\begin{equation}
=\alpha_0 \langle \textbf{M}\textbf{G}_0, \; {\cal L}\Psi \rangle\left(
\sum_{k =1}^m \frac{\Lambda_k |c_k|^2}{\Lambda_k-\Lambda_M} + \int_I
\frac{\Lambda |c(\Lambda)|^2}{\Lambda-\Lambda_M}d\Lambda \right),
\end{equation}
thus we get
\begin{equation}
\langle \textbf{M}\textbf{G}_0, \; {\cal L}\Psi \rangle \cdot
Q(\Lambda_M)=0,
\end{equation}
where
\begin{equation} \label{Q}
Q(\Lambda_M) \equiv \sum_{k=1}^m \frac{\Lambda_k
|c_k|^2}{\Lambda_k-\Lambda_M} + \int_I \frac{\Lambda
|c(\Lambda)|^2}{\Lambda-\Lambda_M}d\Lambda - \frac{1}{\alpha_0}.
\end{equation}
Then the discrete eigenmodes of ${\cal L}_M$ are such that either
\begin{equation} \label{condition1}
\langle \textbf{M}\textbf{G}_0, \; {\cal L}\Psi \rangle=0,
\end{equation}
or
\begin{equation} \label{condition2}
Q(\Lambda_M)=0.
\end{equation}
The continuous eigenvalues of ${\cal L}_M$ are the same as those of
${\cal L}$ since ${\cal L}_M \to {\cal L}$ as $|\textbf{x}| \to
\infty$.

We first consider eigenvalues of ${\cal L}_M$ where condition
(\ref{condition1}) holds. In this case, Eq. (\ref{calLmPsi}) becomes
the eigenvalue equation for ${\cal L}$, thus these eigenvalues of
${\cal L}_M$ are also the eigenvalues of ${\cal L}$. As a result,
\begin{equation} \label{inequal1}
\Lambda_{min} \le \Lambda_M \le 0.
\end{equation}

Next, we consider eigenvalues of ${\cal L}_M$ which satisfy
condition (\ref{condition2}). For these eigenvalues, we will show
that the following inequality holds:
\begin{equation} \label{inequal2}
\min\left(\Lambda_{min}, \bar{\alpha}_0+\Lambda_{min}\right) \le
\Lambda_M <0,
\end{equation}
where
\begin{equation} \label{alphabar}
\bar{\alpha}_0\equiv -\alpha_0\langle \textbf{M}\textbf{G}_0, {\cal
L}\textbf{G}_0\rangle = -\frac{\langle \textbf{M}\textbf{G}_0, \;
{\cal L}\textbf{G}_0 \rangle}{\langle \textbf{MG}_0,
\textbf{G}_0\rangle}-\frac{1}{\Delta t}.
\end{equation}
To prove the right half of this inequality, recall that $\Lambda_k
\le 0$ and $\Lambda \in I <0$, hence for any real number $\Lambda_M
\ge 0$,
\begin{equation} \label{Q1}
0< \sum_{k=1}^m \frac{\Lambda_k |c_k|^2}{\Lambda_k-\Lambda_M} + \int_I
\frac{\Lambda |c(\Lambda)|^2}{\Lambda-\Lambda_M}d\Lambda \le
\sum_{k=1}^m |c_k|^2 +\int_I |c(\Lambda)|^2 d\Lambda = \langle
\textbf{MG}_0, \textbf{G}_0\rangle.
\end{equation}
On the other hand,
\begin{equation}
\frac{1}{\alpha_0}=\frac{\langle \textbf{MG}_0,
\textbf{G}_0\rangle}
{1+\frac{\langle \textbf{MG}_0, \textbf{G}_0\rangle}{\langle
\textbf{M}\textbf{G}_0, {\cal L}\textbf{G}_0\rangle \Delta t }}.
\end{equation}
Since $\textbf{L}_1\textbf{G}_0\ne 0$ by assumption and $\textbf{M}$
is Hermitian and positive definite, then $\langle
\textbf{M}\textbf{G}_0, {\cal L}\textbf{G}_0\rangle<0$. In addition,
$\langle \textbf{MG}_0, \textbf{G}_0\rangle >0$. Thus
\begin{equation} \label{Q2}
\frac{1}{\alpha_0} > \langle \textbf{MG}_0, \textbf{G}_0\rangle
\hspace{0.3cm}\mbox{or} \hspace{0.3cm} \frac{1}{\alpha_0} <0.
\end{equation}
In view of Eqs. (\ref{Q}), (\ref{Q1}) and (\ref{Q2}), $Q(\Lambda_M)$
can only have negative roots. Thus, the right half of inequality
(\ref{inequal2}) holds.

In the following, we prove the left half of inequality
(\ref{inequal2}). Notice that
\begin{equation}
\Lambda_{min}=\min_{\psi}\frac{\langle \textbf{M}\psi, \; {\cal L}\psi
\rangle}{\langle \textbf{M}\psi, \psi\rangle},
\end{equation}
thus $\bar{\alpha}_0+\Lambda_{min}<0$ in view of Eq.
(\ref{alphabar}) and $\Delta t >0$. Let us rewrite $1/\alpha_0$ as
\begin{equation}
\frac{1}{\alpha_0}
=-\frac{\langle \textbf{M}\textbf{G}_0, \; {\cal L}\textbf{G}_0
\rangle}{\bar{\alpha}_0}=-\frac{1}{\bar{\alpha}_0}\left(\sum_{k=1}^m
\Lambda_k |c_k|^2 + \int_I \Lambda |c(\Lambda)|^2 d\Lambda\right).
\end{equation}
When this expression is substituted into Eq. (\ref{Q}), we get
\begin{equation} \label{Qs}
Q(\Lambda_M)=\frac{1}{\bar{\alpha}_0}\left[ \sum_{k=1}^m
\frac{(\Lambda_k+\bar{\alpha}_0-\Lambda_M)\Lambda_k
|c_k|^2}{\Lambda_k-\Lambda_M} + \int_I
\frac{(\Lambda+\bar{\alpha}_0-\Lambda_M)\Lambda
|c(\Lambda)|^2}{\Lambda-\Lambda_M}d\Lambda\right].
\end{equation}
When $\Lambda_M < \min\left(\Lambda_{min},
\bar{\alpha}_0+\Lambda_{min}\right)$, all terms inside the square
bracket of Eq. (\ref{Qs}) are negative, thus $Q(\Lambda_M)$ can not
be zero. As a result, the left half of inequality (\ref{inequal2})
holds.

Due to the two inequalities (\ref{inequal1}) and (\ref{inequal2}) on
eigenvalues $\Lambda_M$ for the two cases (\ref{condition1}) and
(\ref{condition2}), we see that all eigenvalues of ${\cal L}_M$ are
non-positive, and the kernel of ${\cal L}_M$ is the same as the
kernel of $\textbf{L}_1$. In addition, the convergence condition
$|1+\Lambda_M\Delta t|<1$ for non-zero eigenvalues $\Lambda_M$ of
${\cal L}_M$ will be satisfied if
\begin{equation} \label{fir}
\Delta t < -\frac{2}{\Lambda_{min}},
\end{equation}
and
\begin{equation} \label{sec}
\Delta t < -\frac{2}{\bar{\alpha}_0+\Lambda_{min}}.
\end{equation}
Since $\bar{\alpha}_0+\Lambda_{min}<0$ and due to Eq.
(\ref{alphabar}),  the second inequality (\ref{sec}) is equivalent
to
\begin{equation}
\Delta t <\frac{1}{\frac{\langle \textbf{M}\textbf{G}_0, \; {\cal L}\textbf{G}_0
\rangle}{\langle \textbf{MG}_0, \textbf{G}_0\rangle}-\Lambda_{min}}.
\end{equation}
Together with Eq. (\ref{fir}), we find that when $\Delta t$
satisfies the restriction (\ref{dtm}), $|1+\Lambda_M\Delta t|<1$ for
all non-zero eigenvalues $\Lambda_M$ of ${\cal L}_M$.

Lastly, using similar techniques as employed above, we can readily
show that ${\cal L}_M$ has no generalized eigenfunctions at zero
eigenvalue, i.e., equation
\begin{equation}
{\cal L}_M \textbf{F} = \textbf{u}_{c_j}, \hspace{0.5cm} 1\le j \le
s
\end{equation}
has no square-integrable solutions $\textbf{F}$. Here
$\textbf{u}_{c_j}$ is in the kernel of ${\cal L}_M$, i.e. the kernel
of $\textbf{L}_1$ (see above and Assumption 1). This rules out the
possibility of linear (secular) growth of zero-eigenmodes
$\textbf{u}_{c_j}$ in the iteration of the error function
$\tilde{\textbf{u}}_n$. This concludes the proof of Theorem
\ref{theorem2b}.

To illustrate the faster convergence of MSOM, we apply MSOM with the
choice of (\ref{choice1}) to the NLS equation (\ref{NLS}), where
explicit convergence analysis can be carried out.

{\bf Example 3.1} \  Consider MSOM (\ref{MSOM}), (\ref{choice1})
applied to the solitary wave (\ref{soliton}) in the NLS equation
(\ref{NLS}). For simplicity, we take $M$ as in (\ref{M_form}) with
$c=\mu=1$. In this case, by inserting $c=1$ into Eq.
(\ref{Lambdaeq}), we find that the eigenvalues of operator
$M^{-1}L_1$ are
\begin{equation}
\lambda_j=\frac{24}{(2j+3)^2-1}-1,  \hspace{0.5cm} j=0, 1, 2, \dots
\end{equation}
Hence $\lambda_0=2$, $\lambda_1=0$, $\lambda_2=-0.5$, \dots,
$\lambda_\infty=-1$. Eigenvalues of operator ${\cal L}$ are
\begin{equation}
\Lambda=-\lambda_j^2,\hspace{0.5cm} j=0, 1, 2, \dots
\end{equation}
Notice that $M^{-1}L_1u=2u$, hence $G_0=u$ is an eigenfunction of
${\cal L}$. According to the discussions at the beginning of this
section, we find that eigenvalues of ${\cal L}_M$ are
\begin{equation}
\Lambda_{M, 0}=-\frac{1}{\Delta t},
\end{equation}
\begin{equation}
\Lambda_{M, j}=-\lambda_j^2,  \hspace{0.5cm} j=1, 2, \dots
\end{equation}
Hence the convergence factor $R_M$ for MSOM (\ref{MSOM}) is
\begin{equation} \label{RM}
R_M(\Delta t)=\max\left\{|1-\lambda_2^2\Delta t|,
|1-\lambda_\infty^2\Delta t|\right\}=\max\left\{|1-0.25\Delta t|,
|1-\Delta t|\right\}.
\end{equation}
It is noted that the zero eigenvalue $\Lambda_{M, 1}$ corresponds to
a translational-invariance mode and does not affect the convergence
analysis. From Eq. (\ref{RM}), we see that MSOM (\ref{MSOM})
converges if $\Delta t < \Delta t_{max}=2$. The fastest convergence
is achieved at
\begin{equation}
\Delta t=\Delta t_*=\frac{2}{\lambda_2^2+\lambda_\infty^2}=1.6,
\end{equation} and the corresponding convergence factor is
\begin{equation} \label{RM*}
R_{M*}=\frac{\lambda_\infty^2-\lambda_2^2}{\lambda_\infty^2+\lambda_2^2}=0.6.
\end{equation}
This convergence factor is substantially lower than that of SOM
(\ref{SOM}) (see Fig. 1), thus MSOM (\ref{MSOM}), (\ref{choice1})
converges much faster. We note in passing that this convergence
factor (\ref{RM*}) for MSOM is close to those of the optimal
Petviashvili method and the optimally accelerated imaginary-time
evolution method, which can be calculated to be 0.5 and 0.51
respectively \cite{YangLakoba}.

{\bf Remark 3.1} \ Unlike $\Delta t_{max}$ in Theorem \ref{theorem0}
for SOM, $\Delta t_M$ in the time-step restriction (\ref{dtm}) is
not a sharp bound for convergence of MSOM (\ref{MSOM}). In practice,
$\Delta t$ can often be taken larger than $\Delta t_M$, and MSOM
still converges. This can be seen clearly in Example 3.1, where
$\Lambda_{min}=-4$, $G_0=u$, $\langle MG_0, {\cal L}G_0
\rangle/\langle MG_0, G_0\rangle=4$, hence $\Delta t_M=\frac{1}{8}$.
But according to the above explicit calculations, the sharp bound on
the time step is $\Delta t_{max}=2$, much larger than $\Delta t_M$.
Thus, restriction (\ref{dtm}) is sufficient, but not necessary, for
the convergence of MSOM (\ref{MSOM}).

Of the two simple choices for $\textbf{G}_n$ given in
(\ref{choice1}) and (\ref{choice2}), $\textbf{G}_0=\textbf{u}$ for
the first choice (\ref{choice1}), and $\textbf{G}_0=0$ for the
second choice (\ref{choice2}). Thus Theorem \ref{theorem2b} applies
to choice (\ref{choice1}), but not to choice (\ref{choice2}) since
$\textbf{L}_1\textbf{G}_0\equiv 0$ there. The convergence analysis
for MSOM with choice (\ref{choice2}) is more complicated since the
corresponding linearized iteration equation for the error is no
longer a one-step iteration, but rather a two-step iteration.
However, choice (\ref{choice2}) has certain intuitive appeal, as the
difference function $\textbf{u}_n-\textbf{u}_{n-1}$ often resembles
the slowest decaying mode in the error \cite{LY06}, thus it makes
sense to choose (\ref{choice2}). Our numerical experiments have
shown that indeed, MSOM with (\ref{choice2}) not only converges
(when the time step is below a certain threshold), but also
converges much faster than the choice (\ref{choice1}) in almost all
cases. In fact, we will see from various examples in Sec. 6 that
MSOM (\ref{MSOM}) with choice (\ref{choice2}) often gives the
fastest convergence among schemes proposed in this paper, especially
for solitary waves with complicated profiles.

\section{Power-conserving and arbitrary-quantity-conserving squared-operator iteration methods}

In some applications, solitary waves are sought with a pre-specified
power (or other quantity such as energy) rather than the propagation
constant. For instance, in Bose-Einstein condensation, the number of
atoms is often specified, and a solitary wave with that number of
atoms (i.e. power in our notations) needs to be computed. In
principle, we can still use the aforementioned SOM or MSOM to
compute the wave by first continuously varying propagation constants
to obtain the power curve, then determining the propagation constant
with the pre-specified power, then finally computing the wave again
with SOM/MSOM. But that is clearly awkward and not time efficient. A
much better way is to design numerical schemes which automatically
converge to the solution with the pre-specified quantity. Another
example is the linear eigenvalue problem. In this case, the
eigenvalues are unknown, thus SOM/MSOM clearly does not apply. If
one views the eigenvalues as propagation constants, and requires the
eigenfunctions to have a fixed power(norm), then this linear
eigenvalue problem becomes the same as a solitary-wave problem with
a pre-specified power. To treat this type of problems, one can use
the imaginary-time evolution method
\cite{GarciaRipollP_01,Bao_Du,Bao_04,VS_04,YangLakoba}, where the
solution is normalized at every step to keep the pre-specified
power. However, the problem with the imaginary-time evolution method
is that it often diverges when the solution crosses zero
\cite{YangLakoba}. Thus new numerical schemes which conserve the
power but also have guaranteed convergence need to be constructed.

In this section, we propose power- (or any other quantity-)
conserving squared operator methods, where one specifies the power
(or any other quantity) instead of the propagation constant. These
methods are in the same spirit as the one presented in
\cite{GarciaRipollP_01}, but we go much further here. First, we will
rigorously prove that these new power-conserving methods converge to
any solitary wave as long as the time step is below a certain
threshold, and the initial condition is close to that solution ---
just like SOM/MSOM. This has never been done before. We should point
out that this guaranteed convergence for all solitary waves is
nontrivial; it is certainly not automatic for an iteration method
even if the "squared operator" idea has been incorporated. This
guaranteed convergence is possible only when the updating formula
for the propagation constants is compatible with the solution
updating scheme, in addition to the "squared operator" idea. Second,
the methods we will propose use a different acceleration than the
acceleration of \cite{GarciaRipollP_01}, and hence, as we show in
the Appendix, have faster convergence. Thirdly, our methods apply to
all types of equations and arbitrary forms of conserved quantities,
more general than the method of Ref. \cite{GarciaRipollP_01}.

For the ease of presentation, we first consider three special (yet
large) classes of equations and construct their power-conserving
squared-operator methods (PCSOM). Then we consider the most general
case where the wave equations and the pre-specified quantities of
the solution are both arbitrary, and construct their
quantity-conserving squared-operator methods (QCSOM).

\subsection{The power-conserving squared-operator method for equations with a single propagation constant}
Consider equation (\ref{L0U}) of the form
\begin{equation} \label{L00u}
{\bf L}_0{\bf u} \equiv {\bf L}_{00}{\bf u} - \mu {\bf u} =0.
\end{equation}
Here ${\bf u}(\x; \mu)$ is a real vector solitary wave, and $\mu$ is
a real scalar constant (usually called the propagation constant).
Solution ${\bf u}$ is assumed to exist for continuous ranges of
$\mu$ values. Equations of the form (\ref{L00u}) include all scalar
equations as well as certain vector equations such as vortex
equations (\ref{NLS2Du})-(\ref{NLS2Dv}) in Example 6.1 and the
second-harmonic generation system (\ref{SHG1})-(\ref{SHG2}) in
Example 6.2. Define the power of solution ${\bf u}(\x; \mu)$ as
\begin{equation}
P(\mu)=\langle {\bf u}(\x; \mu),  {\bf u}(\x; \mu) \rangle,
\end{equation}
then we are interested in finding a solution ${\bf u}(\x; \mu)$
whose power has a pre-specified value $P$. Combining ideas of SOM
(\ref{SOM}) and the imaginary-time evolution method
\cite{GarciaRipollP_01,Bao_Du,YangLakoba}, we propose the following
new power-conserving squared-operator method (PCSOM):
\begin{equation} \label{PCSOM}
{\bf u}_{n+1}=\left[\frac{P}{\langle \hat{\bf u}_{n+1}, \hat{\bf u}_{n+1}
\rangle}\right]^{\frac{1}{2}} \hat{\bf u}_{n+1},
\end{equation}
where
\begin{equation} \label{PCSOMb}
\hat{\bf u}_{n+1}={\bf u}_n- {\bf M}^{-1}\left[{\bf L}_1^\dagger {\bf M}^{-1}{\bf L}_0{\bf u}-\gamma
{\bf u}\right]_{{\bf u}={\bf u}_n, \: \mu=\mu_n} \Delta t,
\end{equation}
${\bf M}$ is a positive-definite and Hermitian acceleration operator, and
\begin{equation} \label{mun2}
\gamma=\frac{\langle {\bf u}, \; {\bf M}^{-1}{\bf L}_1^\dagger {\bf M}^{-1}{\bf L}_0{\bf u} \rangle
}{\langle {\bf u}, \; {\bf M}^{-1}{\bf u} \rangle }, \hspace{0.5cm}
\mu_n=\left. \frac{\langle {\bf u}, \; {\bf M}^{-1}{\bf L}_{00}{\bf u} \rangle}{\langle {\bf u}, \;
{\bf M}^{-1}{\bf u} \rangle }\right|_{{\bf u}={\bf u}_n}.
\end{equation}
%
The convergence property of this PCSOM is similar to that of the SOM, and is
summarized in the following theorem.
\begin{theorem} \label{AACITtheorem}
Let Assumption 1 be valid, ${\bf u}(\x)$ be orthogonal to the kernel
of ${\bf L}_1$, $P'(\mu) \equiv dP(\mu)/d\mu \ne 0$, and define
$\Delta t_{max}$ by Eq. (\ref{dtmax}), where $\Lambda_{min}$ now is
the minimum eigenvalue of the iteration operator ${\cal L}_{PC}$
defined in Eq. (\ref{hatLA}) below. Then when $\Delta t < \Delta
t_{max}$, PCSOM (\ref{PCSOM})-(\ref{mun2}) is guaranteed to converge
to the solitary wave ${\bf u}(\x)$ of Eq. (\ref{L00u}) with power
$P$ if the initial condition is close to ${\bf u}(\x)$. When $\Delta
t > \Delta t_{max}$, PCSOM (\ref{PCSOM})-(\ref{mun2}) diverges.
\end{theorem}

{\bf Proof.}  \ As before, we use the linearization method to
analyze PCSOM (\ref{PCSOM})-(\ref{mun2}). Substituting Eq.
(\ref{linearize}) into the scheme and linearizing, we find that the
iteration equation for the error $\tilde{\bf u}_n$ is
\begin{equation} \label{errorun}
\tilde{\bf u}_{n+1}=\tilde{\bf u}_n + {\cal L}_{PC} \: \tilde{\bf u}_n \: \Delta t,
\end{equation}
where operator ${\cal L}_{PC}$ is
\begin{equation} \label{hatLA}
{\cal L}_{PC} \Psi ={\bf M}^{-1}\left( \hat{\bf L} \Psi - \hat{\gamma} {\bf u}
\right),
\end{equation}
\begin{equation}
\hat{\bf L} \Psi=-{\bf L}_1^\dagger {\bf M}^{-1}\left({\bf L}_1\Psi-\beta {\bf u}
\right),
\end{equation}
and
\begin{equation} \label{gammabeta}
\hat{\gamma}=\frac{\langle {\bf u}, \; {\bf M}^{-1}\hat{\bf L}\Psi \rangle }{\langle {\bf u}, \;
{\bf M}^{-1}{\bf u}   \rangle }, \hspace{0.5cm}
\beta = \frac{\langle {\bf u}, \; {\bf M}^{-1}{\bf L}_1\Psi \rangle }{\langle {\bf u}, \; {\bf M}^{-1}{\bf u}
\rangle}.
\end{equation}
In addition, due to the power normalization (\ref{PCSOM}), we also
have the constraint:
\begin{equation} \label{constraint}
\langle \tilde{\bf u}_n,  {\bf u} \rangle =0.
\end{equation}
Thus it is sufficient to consider eigenfunctions $\Psi(\x)$ of operator ${\cal L}_{PC}$ in the
restricted space $S$:
\begin{equation}\label{S}
S\equiv \{ \Psi(\x): \ \langle \Psi, {\bf u} \rangle =0\}.
\end{equation}

First, we rewrite the operator ${\cal L}_{PC}$ as ${\cal
L}_{PC}={\bf M}^{-1/2}{\cal L}_{PCH}{\bf M}^{1/2}$, which defined
operator ${\cal L}_{PCH}$. Then ${\cal L}_{PC}$ and ${\cal L}_{PCH}$
are similar, hence having the same eigenvalues. Since we only need
to consider eigenfunctions of ${\cal L}_{PC}$ which are orthogonal
to ${\bf u}$, it follows that we only need to consider
eigenfunctions of ${\cal L}_{PCH}$ which are orthogonal to ${\bf
M}^{-1/2}{\bf u}$. In this space, it is easy to check that operator
${\cal L}_{PCH}$ is Hermitian. In addition, using the
Cauchy-Schwartz inequality, we can verify that ${\cal L}_{PCH}$ is
also semi-negative definite. Thus, all eigenvalues of ${\cal
L}_{PC}$ are real and non-positive, and the eigenfunctions of ${\cal
L}_{PC}$ form a complete set in space $S$. Another way to prove
these results is to notice that the operator $\hat{\bf L}$ is
Hermitian and semi-negative definite by inspection and using the
Cauchy-Schwartz inequality. Thus operator $\hat{\bf L} \Psi -
\hat{\gamma} {\bf u}$ is Hermitian and semi-negative definite in the
space $S$. Hence according to the Sylvester inertia law (see, e.g.,
Theorems 4.5.8 and 7.6.3 in \cite{HornJohnson91}), all eigenvalues
of ${\cal L}_{PC}$ are real and non-positive, and the eigenfunctions
of ${\cal L}_{PC}$ form a complete set in space $S$.
As a result, under the time-step restriction $\Delta t
< \Delta t_{max}$,  the convergence condition $|1+\Lambda_{PC}\Delta
t|<1$ holds for all non-zero eigenvalues $\Lambda_{PC}$ of ${\cal
L}_{PC}$. On the other hand, if $\Delta t > \Delta t_{max}$,
$1+\Lambda_{min}\Delta t < -1$, hence iterations diverge.

To complete the proof of Theorem 3, it remains to consider the kernel of ${\cal L}_{PC}$ in space $S$ and verify
that functions $\Psi$ in this kernel do not affect the convergence of the iterations. For these functions,
we have
\begin{equation}
\hat{\bf L}\Psi- \hat{\gamma} {\bf u}=0.
\end{equation}
Taking the inner product between this equation and $\Psi$, and
noticing $\Psi \in S$, we get
\begin{equation}
\langle \hat{\bf L}\Psi, \Psi \rangle=0.
\end{equation}
Since $\hat{\bf L}$ is semi-negative definite, from the
Cauchy-Schwartz inequality and the condition for its equality to
hold, we get
\begin{equation} \label{L1Psi}
{\bf L}_1\Psi = \beta {\bf u}.
\end{equation}
On the other hand, differentiating Eq. (\ref{L00u}) with respect to
$\mu$, we find that
\begin{equation}
{\bf L}_1 {\bf u}_\mu ={\bf u}.
\end{equation}
Thus the solution $\Psi$ of Eq. (\ref{L1Psi}) is equal to $\beta
{\bf u}_{\mu}$ plus functions in the kernel of ${\bf L}_1$. Due to
the constraint $\Psi \in S$ and recalling our assumptions of
$\langle \textbf{u}_{\mu},\textbf{u} \rangle = \frac12 P'(\mu) \neq
0$ and $\textbf{u}$ being orthogonal to the kernel of
$\textbf{L}_1$, we get $\beta=0$. Thus the kernel of ${\cal L}_{PC}$
in space $S$ is the same as the kernel of ${\bf L}_1$. Since this
kernel only contains invariance modes by Assumption 1, it does not
affect convergence of the iterations.
%
%
Combining all the results obtained above, Theorem \ref{AACITtheorem} is then proved.

\subsection{The power-conserving squared-operator method for $K$ equations with $K$ propagation constants}

Next, we consider another class of equations (\ref{L0U}) in the form
\begin{equation} \label{L0U_vector}
{\bf L_{0} u} \equiv {\bf L_{00} u} - {\rm diag}(\mu_1, \ldots, \mu_K) \, {\bf u}
= \bf 0,
\end{equation}
where $\textbf{u}(\x)=[u_1, u_2, \dots, u_K]^T$ is a real vector
solitary wave, the superscript "T" represents the transpose of a
vector, and $\mu_k$'s are real propagation constants which are
assumed to be independent of each other. The key feature of this
case is that the number of independent propagation constants $\mu_k$
is equal to the number of components in the vector solution
$\textbf{u}(\x)$.
Defining the powers of individual components $u_k(\x)$ of the
solution as
\begin{equation}
\label{P_k} P_k\equiv \langle u_k, u_k \rangle, \hspace{0.5cm} 1\le
k \le K,
\end{equation}
and introducing the following $K\times K$ matrix
\begin{equation} \label{D}
\textbf{D}_{K\times K} \equiv \left(\frac{\partial
P_i}{\partial\mu_j} \right),
\end{equation}
then the PCSOM for Eq. (\ref{L0U_vector}) we propose is
\begin{equation} \label{PCSOM_vector}
u_{k, n+1}=\left[\frac{P_k}{\langle \hat{u}_{k, n+1}, \hat{u}_{k,
n+1} \rangle}\right]^{\frac{1}{2}} \hat{u}_{k, n+1}, \hspace{0.5cm}
1\le k \le K,
\end{equation}
where
\begin{equation} \label{PCSOM2_vector}
\hat{\textbf{u}}_{n+1}=\textbf{u}_n-
{\mathbf{M}}^{-1}\left\{{\mathbf{L}}_1^\dagger{\mathbf{M}}^{-1}{\mathbf{L}}_0\textbf{u}-\mbox{diag}(\gamma_1,
\dots, \gamma_K) \textbf{u}\right\}_{\textbf{u}=\textbf{u}_n, \
\mu_k=\mu_{k, n}}\Delta t,
\end{equation}
\begin{equation} \label{PCSOM3_vector}
\gamma_k =\frac{\langle u_k, \;
[{\mathbf{M}}^{-1}{\mathbf{L}}_1^\dagger{\mathbf{M}}^{-1}{\mathbf{L}}_0\textbf{u}]_k
\rangle}{\langle u_k, \; M_k^{-1}u_k \rangle}, \hspace{0.5cm}
\mu_{k, n}=\left. \frac{ \langle u_k, \; [{\bf M^{-1}L_{00} u}]_k  \rangle}{\langle u_k, \; M_k^{-1}u_k
 \rangle} \right|_{\textbf{u}=\textbf{u}_n}.
\end{equation}
Here $[\cdot]_k$ represents the $k$-th component of the vector
inside the bracket, $[\cdot]_{k, n}$ represents the $n$-th iteration
of the $k$-th component of the vector inside the bracket, and ${\bf
M}\equiv \mbox{diag}(M_1, M_2, \dots, M_K)$ is a positive definite
and Hermitian acceleration operator.

The convergence properties of this PCSOM for Eq. (\ref{L0U_vector})
are similar to those of the previous PCSOM, and they are summarized
in the following theorem.
\begin{theorem} \label{theorem2_vector}
Let all $\mu_k$'s be independent, Assumption 1 be valid, ${\bf
u}(\x)$ be orthogonal to the kernel of ${\bf L_1}$,
$\mbox{det}(\textbf{D}) \ne 0$, and define $\Delta t_{max}$ by Eq.
(\ref{dtmax}), where $\Lambda_{min}$ here is the minimum eigenvalue
of the operator ${\cal L}_{PCV}$ defined as
\begin{equation} \label{calLhat2_vector}
{\cal L}_{PCV}\Psi = {\mathbf{M}}^{-1}\left\{\hat{\mathbf{L}}_V\Psi-\mbox{diag}
(\hat{\gamma_1}, \dots, \hat{\gamma_K})\textbf{u}\right\},
\end{equation}
where
\begin{equation}
\hat{\mathbf{L}}_V\Psi=-{\mathbf L}_1^\dagger{\mathbf
M}^{-1}\left[{\mathbf L}_1\Psi- \mbox{diag}\left(\beta_1, \dots,
\beta_K \right) \textbf{u}\right],
\end{equation}
and
\begin{equation}
\hat{\gamma_k}=\frac{\langle u_k, \;
[{\mathbf{M}}^{-1}\hat{\mathbf{L}}_V\Psi]_k  \rangle }{\langle u_k, \;
M_k^{-1}u_k  \rangle },  \hspace{0.5cm}
\beta_k=\frac{\langle u_k, \; [{\mathbf{M}}^{-1}{\mathbf L}_1\Psi]_k
\rangle}{\langle u_k, \; M_k^{-1}u_k  \rangle}.
\end{equation}
Then when $\Delta t < \Delta t_{max}$,
PCSOM (\ref{PCSOM_vector})-(\ref{PCSOM3_vector}) is guaranteed
to converge to the solitary wave $\textbf{u}(\x)$ of Eq. (\ref{L0U_vector}) with component powers
being $[P_1, \dots, P_k]^T$, if the initial condition is close to $\textbf{u}(\x)$.
When $\Delta t > \Delta t_{max}$, PCSOM (\ref{PCSOM_vector})-(\ref{PCSOM3_vector}) diverges.
\end{theorem}

{\bf Proof.} The proof of this theorem is analogous to that of Theorem \ref{AACITtheorem}.
By the linearization analysis, we find that the error of the iterated function satisfies the equation
\begin{equation}
\tilde{\bf u}_{n+1}=\tilde{\bf u}_n + {\cal L}_{PCV} \: \tilde{\bf u}_n \: \Delta
t,
\end{equation}
where operator ${\cal L}_{PCV}$ is as defined in Eq. (\ref{calLhat2_vector}).
The power normalization (\ref{PCSOM_vector}) implies that it suffices to consider the
eigenfunctions $\Psi$ of ${\cal L}_{PCV}$ satisfying the following orthogonality
conditions:
\begin{equation} \label{orth_vec}
\langle \Psi_k, u_k \rangle
=0, \hspace{0.5cm} k=1, \dots, K.
\end{equation}
Similar to the operator ${\cal L}_{PC}$ in the proof of Theorem
\ref{AACITtheorem}, we can show that in the space of functions
satisfying the above orthogonality conditions, all eigenvalues of
${\cal L}_{PCV}$ are real and non-positive, and all its
eigenfunctions form a complete set. The main difference between the
previous operator ${\cal L}_{PC}$ and ${\cal L}_{PCV}$ here is in
their kernels. The kernel of operator ${\cal L}_{PCV}$ contains
functions $\Psi$ where ${\cal L}_{PCV}\Psi =0$.
%
Similar to the case in Theorem \ref{AACITtheorem}, we can show that
\begin{equation} \label{L1Psi_vec}
{\mathbf{L}}_1\Psi=\mbox{diag}(\beta_1, \beta_2, \dots, \beta_K)
\textbf{u}.
\end{equation}
Differentiating Eq. (\ref{L0U_vector}) with respect to $\mu_k$, we get
\begin{equation}
 {\mathbf{L}}_1\textbf{u}_{\mu_k}=(0, \ldots, u_k, \ldots, 0)^T, \hspace{1cm} k=1,
\dots, K.
\end{equation}
Hence the solution of Eq. (\ref{L1Psi_vec}) is
\begin{equation} \label{Psiform}
\Psi=\sum_{k=1}^K \beta_k\textbf{u}_{\mu_k}
\end{equation}
plus the homogeneous solution of Eq. (\ref{L1Psi_vec}).
Substituting this equation into the orthogonality conditions
(\ref{orth_vec}), and recalling the assumptions of Theorem \ref{theorem2_vector},
we get
\begin{equation} \label{DC}
\textbf{D}\cdot \vec{\beta}=0,
\end{equation}
where matrix $\textbf{D}$ is defined in Eq. (\ref{D}), and
$\vec{\beta}=(\beta_1, \beta_2, \dots, \beta_K)^T$.  Since
$\mbox{det}(\textbf{D})\ne 0$ by assumption, the solution to Eq.
(\ref{DC}) is $\vec{\beta}=\vec{0}$. Hence the kernel of ${\cal
L}_{PCV}$ satisfying the orthogonality conditions (\ref{orth_vec})
is the same as the kernel of ${\mathbf L}_1$. Thus, Theorem
\ref{theorem2_vector} is proved.

The PCSOMs (\ref{PCSOM})--(\ref{mun2}) and
(\ref{PCSOM_vector})-(\ref{PCSOM3_vector}) have been presented above
as separate cases because they have simple forms and also apply to
many equations that arise frequently in applications. For example,
the method of Section 4.1 applies to linearly coupled nonlinear
Schr\"odinger equations, while the method of Section 4.2 applies to
nonlinearly coupled ones; see also Examples 6.1--6.3 below. A PCSOM
would also result for a more general case where in Eq. (\ref{L0U})
of the form (\ref{L0U_vector}), the propagation constants $\mu_k$'s
can be separated into several independent groups of equal $\mu_k$'s.
An example of such a system of equations can be found, e.g., in
\cite{LakobaK_97}. In this case, the PCSOM can be constructed by
combining the two PCSOMs (\ref{PCSOM})-(\ref{mun2}) and
(\ref{PCSOM_vector})-(\ref{PCSOM3_vector}). Here we group together
the solution components $u_k$'s whose corresponding propagation
constants $\mu_k$'s are the same. These groups form sub-vectors in
the whole solution vector ${\bf u}(\x)$. Then the PCSOM for this
more general equation is analogous to PCSOM
(\ref{PCSOM_vector})-(\ref{PCSOM3_vector}), except that $u_k$ is
replaced by each sub-vector of the solution, and $P_k$ replaced by
that sub-vector's power. The convergence properties of this more
general PCSOM are similar to those in Theorems \ref{AACITtheorem}
and \ref{theorem2_vector}, and will not be elaborated here.

\subsection{The squared-operator method with general quadratic conserved quantities}

In this subsection, we present the PCSOM for the case where the
conserved quantities are not restricted to simple powers (as in
Secs. 4.1 and 4.2). Rather, they can be more general quadratic
quantities of the solutions (i.e. linear combinations of powers of
individual components of the vector solitary wave). This case
includes as particular cases the PCSOMs presented in Sections 4.1
and 4.2. We chose to present it separately from those two methods
because its form is more complex than theirs, while the situations
where neither of those two simpler methods could apply are
relatively rare. One example of such a situation is the system of
coherent three-wave interactions in a quadratic medium
\cite{3w1,3w2}, which has only two conserved quantities which are
linear combinations of the three wave powers (these conservation
laws are known as Manley-Rowe relations).

The class of problems described above can be formulated
mathematically as follows:
\begin{equation} \label{L0U_quadr}
{\bf L}_0{\bf u} \equiv {\bf L}_{00}{\bf u} -\frac{\delta {\bf
Q}}{\delta {\bf u}} \vec{\mu} =0,
\end{equation}
and
\begin{equation} \label{Qconstraint_quadr}
Q_j({\bf u})\equiv \sum_{l=1}^K q_{jk}P_k=C_j,   \hspace{0.5cm} j=1,
\dots, r.
\end{equation}
Here $\textbf{u}(\x)=[u_1, u_2, \dots, u_K]^T$ is a vector solitary
wave, $\delta/\delta {\bf u}=[\delta/\delta u_1, \ldots,
\delta/\delta u_K]^T$ is the functional derivative,
$\vec{\mu}=(\mu_1, \dots, \mu_r)^T$ is the vector of $r$ {\it
linearly independent} propagation constants ($1\le r \le K$), $P_k$
are defined in (\ref{P_k}), $q_{jk}$ are constants, and $C_j$ are
the specified values of the quadratic conserved quantities $Q_j$.
All variables and constants involved are real-valued. Without any
loss of generality, we assume that the $r\times K$ matrix of the
coefficients $q_{jk}$ is in the reduced echelon form such that
\begin{equation}
q_{ij}=\left\{\begin{array}{ll} 0,  & i > j, \\
                            1,  &  i=j.
                            \end{array} \right.
\label{q_jj}
\end{equation}
Then, using the terminology of textbooks in introductory Linear
Algebra, we refer to powers $\{P_k\}_{k=r+1}^K$ and
$\{P_k\}_{k=1}^r$  as "independent" and ``dependent" powers,
respectively.

The PCSOM that we propose for Eqs.
(\ref{L0U_quadr})-(\ref{Qconstraint_quadr}) has the following form:
\be
u_{k,n+1}= \left[
\frac{ C_k - \sum_{l=r+1}^K q_{kl}\hat{P}_l}{ \langle \hat{u}_{k,n+1}, \hat{u}_{k,n+1} \rangle }
 \right]^{1/2} \hat{u}_{k,n+1}, \qquad   1 \le k \le r,
\label{g_normstep}
\ee
\begin{equation} \label{g_PCSOM}
\hat{\bf u}_{n+1}={\bf u}_n- {\bf M}^{-1}\left[{\bf L}_1^\dagger
{\bf M}^{-1}{\bf L}_0{\bf u} - {\bf B} \vec{\gamma}_n \right]_{{\bf
u}={\bf u}_n, \: \vec{\mu}=\vec{\mu}_{n}} \Delta t,
\end{equation}
where the notations for ${\bf L}_1$, ${\bf M}$, and the subscripts
are the same as before, $\hat{P}_k=\langle \hat{u}_{k,n+1} ,
\hat{u}_{k,n+1} \rangle$,
\be
{\bf B} \equiv \frac{\delta {\bf
Q}}{\delta {\bf u}}, \label{B_dQdu} \ee
\be \vec{\gamma}_n= \left. \langle {\bf B}, \; {\bf M}^{-1}{\bf B}
\rangle^{-1} \langle {\bf B}, {\bf M}^{-1}{\bf L}_1^\dagger
\textbf{M}^{-1} {\bf L}_{0}{\bf u} \rangle\right|_{{\bf u}={\bf
u}_n}, \label{vec_gamma_n}
\end{equation}
and
\begin{equation} \label{muQC}
\vec{\mu}_n = \left. \langle {\bf B}, \; {\bf M}^{-1}{\bf B} \rangle^{-1} \langle {\bf B},
{\bf M}^{-1}{\bf L}_{00}{\bf u} \rangle\right|_{{\bf u}={\bf u}_n}.
\end{equation}
The power-normalization step (\ref{g_normstep}) is written with the
account of the reduced echelon form (\ref{q_jj}) of the matrix
$(q_{jk})$. Note that this step affects only the ``dependent"
powers, while the ``independent" ones do not need to be normalized.

Convergence conditions of the PCSOM (\ref{g_normstep})--(\ref{muQC})
are similar to those of the PCSOMs described above. Introducing the
notation
\begin{equation} \label{Dhat}
\hat{\textbf{D}}_{r\times r} \equiv \left(\frac{\partial
Q_i}{\partial\mu_j} \right),
\end{equation}
then convergence conditions of this PCSOM is summarized in the
following theorem.
\begin{theorem} \label{theorem_g_PC}
Let Assumption 1 be valid, all columns of the matrix ${\bf B}$ in
(\ref{B_dQdu}) be orthogonal to the kernel of ${\bf L_1}$, and
$\rm{det}(\hat{\textbf{D}}) \ne 0$. Also, let $\Delta t_{max}$ be
given by Eq. (\ref{dtmax}) where $\Lambda_{min}$ now is the minimum
eigenvalue of the operator ${\cal L}_{PCG}$ defined in Eq.
(\ref{LG}) below. Then for $\Delta t < \Delta t_{max}$, the PCSOM
(\ref{g_normstep})-(\ref{muQC}) is guaranteed to converge to the
solitary wave $\textbf{u}(\x)$ of Eq. (\ref{L0U_quadr}) which
satisfies the constraints (\ref{Qconstraint_quadr}), provided that
the initial condition is close to $\textbf{u}(\x)$. When $\Delta t >
\Delta t_{max}$, the PCSOM (\ref{g_normstep})-(\ref{muQC}) diverges.
\end{theorem}

The proof of this theorem follows the lines of Theorems
\ref{AACITtheorem} and \ref{theorem2_vector}, thus we will only
sketch it below, emphasizing the differences from the proofs of the
latter theorems.

{\bf Proof.} \ Linearizing Eq. (\ref{g_PCSOM}), and noticing that
the power normalization step (\ref{g_normstep}) does not affect the
linearized equations (as in Sections 4.1 and 4.2), we find that the
error satisfies the iteration equation similar to (\ref{errorun}),
except that the iteration operator now becomes
\begin{equation}
\label{LG} {\cal L}_{PCG} \Psi ={\bf M}^{-1}\left( \hat{\bf L}_{PCG}
\Psi - {\bf B} \langle {\bf B}, \; {\bf M}^{-1}{\bf B} \rangle^{-1}
\langle {\bf B}, {\bf M}^{-1} \hat{\bf L}_{PCG} \Psi \rangle
\right),
\end{equation}
where
\begin{equation}
\label{LPCG} \hat{\bf L}_{PCG} \Psi=-{\bf L}_1^\dagger {\bf
M}^{-1}\left({\bf L}_1\Psi- {\bf B}\langle {\bf B}, \;
 {\bf M}^{-1}{\bf B} \rangle^{-1} \langle {\bf B}, \; {\bf M}^{-1}{\bf L}_1 \Psi
 \rangle
\right).
\end{equation}
The eigenfunctions $\Psi$ satisfy the orthogonality relation
\be
\langle B_j, \Psi \rangle = 0, \qquad j = 1, \ldots, r,
\label{orth_g_PC}
\ee
with $B_j$ being the $j$-th column of ${\bf B}$. These conditions
follow from the quantities $Q_j$ being conserved and from the
relation (\ref{B_dQdu}) between ${\bf B}$ and $Q_j$'s.

The operator $\hat{\bf L}_{PCG}$ is Hermitian. In addition, using
the generalized Cauchy-Schwartz inequality for any matrix functions
of ${\bf F}_1$ and ${\bf F}_2$:
\begin{equation} \label{schwartz}
\langle {\bf F}_1, {\bf F}_1 \rangle \ge  \langle {\bf F}_1, {\bf
F}_2 \rangle \langle {\bf F}_2, {\bf F}_2 \rangle^{-1} \langle {\bf
F}_2, {\bf F}_1 \rangle,
\end{equation}
one can verify that $\hat{\bf L}_{PCG}$ is semi-negative definite.
Then following the lines of the proof of Theorem \ref{AACITtheorem},
one then finds that in the space of functions satisfying the
orthogonality conditions (\ref{orth_g_PC}), all eigenvalues of
${\cal L}_{PCG}$ are real and non-positive, and all its
eigenfunctions form a complete set.

Then it remains to consider the kernel of ${\cal L}_{PCG}$, which satisfies the equation
\begin{equation} \label{LPCGker}
\hat{\bf L}_{PCG} \Psi - {\bf B}
\langle {\bf B}, \; {\bf M}^{-1}{\bf B} \rangle^{-1} \langle {\bf B},
{\bf M}^{-1} \hat{\bf L}_{PCG} \Psi \rangle =0,
\end{equation}
and show that the eigenfunctions of this kernel can only be those in
the kernel of ${\bf L}_1$ and thus would not affect the convergence
of this PCSOM in view of our assumptions. To that end, we take the
inner product between Eq. (\ref{LPCGker}) and $\Psi$, use the
orthogonality relations (\ref{orth_g_PC}), then recall that operator
$\hat{\bf L}_{PCG}$ is semi-negative definite, and finally notice
that the Cauchy-Schwartz inequality (\ref{schwartz}) becomes an
equality if and only if ${\bf F}_1$ and ${\bf F}_2$ are linearly
related by a constant matrix. This yields
\begin{equation} \label{L1PsiB}
{\bf L}_1 \Psi = {\bf B}\vec{\beta},
\end{equation}
where $\vec{\beta}=(\beta_1, \dots, \beta_r)^T$ is a constant
vector. Noticing the relations obtained by differentiating Eq.
(\ref{L0U_quadr}) with respect to $\mu_j$ $(1\le j \le r)$, we see,
similarly to (\ref{Psiform}), that the solution to Eq.
(\ref{L1PsiB}) is
\begin{equation}
\Psi=\sum_{j=1}^r \beta_j {\bf u}_{\mu_j}
\end{equation}
plus functions in the kernel of ${\bf L}_1$. Substituting this
solution into (\ref{orth_g_PC}) and recalling that by the
assumptions of Theorem \ref{theorem_g_PC}, columns of ${\bf B}$ are
orthogonal to the kernel of ${\bf L}_1$ and $\mbox{det}(\hat{\bf
D})\neq 0$,  we find that $\vec{\beta}=0$. Thus the kernel of ${\cal
L}_{PCG}$ is the same as that of ${\bf L}_1$. Summarizing all these
results, Theorem \ref{theorem_g_PC} is then proved.

\subsection{The squared-operator method for general equations with arbitrary conserved quantities}

The most general case is probably that the equations depend on the
propagation constants $\mu_k$'s in a general (but linear) way, and
the specified conserved quantities of the solutions are not
restricted to powers or linear combinations of powers, but can be
arbitrary. These equations and constraints can be written in the
general form
\begin{equation} \label{L0UB}
{\bf L}_0{\bf u} \equiv {\bf L}_{00}{\bf u} - {\bf B(u)}\vec{\mu}
=0,
\end{equation}
and
\begin{equation} \label{Qconstraint}
Q_j({\bf u})=C_j,   \hspace{0.5cm} j=1, \dots, r,
\end{equation}
where $\mu=(\mu_1, \dots, \mu_r)^T$ is the vector of all independent
propagation constants, ${\bf B}$ is a general matrix function of
${\bf u}$, and $Q_j({\bf u})$ are arbitrary functionals which are
pre-specified. All quantities involved are real-valued. This system
(\ref{L0UB})-(\ref{Qconstraint}) generalizes the system
(\ref{L0U_quadr})-(\ref{Qconstraint_quadr}) in the previous
subsection in two significant ways. First, the functionals $Q_j$ are
not restricted to the quadratic form (\ref{Qconstraint_quadr}) of
$\textbf{u}$, but are allowed to be arbitrary functionals. For
instance, one can seek a solution with a prescribed value of the
Hamiltonian. Second, matrix $\textbf{B}$ is not restricted to the
special form (\ref{B_dQdu}) as in Eq. (\ref{L0U_quadr}), but is
allowed to be arbitrary functions of $\textbf{u}$. In this general
case, counterparts of the power normalization step
(\ref{g_normstep}) become impossible. Thus, for the solution to have
the pre-specified quantities (\ref{Qconstraint}), new ideas are
necessary. Our idea is to replace the power normalization step with
adding new terms into the iteration step (\ref{g_PCSOM}) in such a
way that when iterations converge, the final solution is guaranteed
to meet the constraints (\ref{Qconstraint}). Our proposed scheme for
this general case with arbitrary conserved quantities, which we
denote QCSOM, is
\begin{equation} \label{QCSOM}
{\bf u}_{n+1}={\bf u}_n- {\bf M}^{-1}\left[{\bf L}_1^\dagger {\bf
M}^{-1}{\bf L}_0{\bf u} + h\: \sum_{j=1}^{r} \left(Q_j({\bf
u})-C_j\right)\frac{\delta Q_j}{\delta {\bf u}} \right]_{{\bf
u}={\bf u}_n, \: \vec{\mu}=\vec{\mu}_n} \Delta t,
\end{equation}
where $\vec{\mu}_n$ is defined by Eq. (\ref{muQC}), and $h>0$ is a
user-specified free scheme parameter whose purpose will be explained
after the proof of Theorem \ref{theorem_QC}. The idea behind this
scheme is that instead of minimizing the functional appearing on the
right-hand side of Eq. (\ref{spanish_functional}) in Section 2, one
minimizes a modified functional which equals the one from Eq.
(\ref{spanish_functional}) {\em plus} additional terms
$\frac{1}{2}\sum[Q_j({\bf u})-C_j]^2$. The acceleration (with the
operator $\textbf{M}^{-1}$) of this scheme is performed in the same
way as the acceleration of (\ref{SOM0}).

On the convergence of this QCSOM for Eqs. (\ref{L0UB}) and (\ref{Qconstraint}), we have the following theorem,
which is very similar to Theorem \ref{theorem_g_PC} of the previous subsection.
\begin{theorem} \label{theorem_QC}
Let Assumption 1 be valid, $\delta Q_j/\delta {\bf u}\;  (j=1, \dots
r)$ be orthogonal to the kernel of ${\bf L_1}$,
$\rm{det}(\hat{\textbf{D}}) \ne 0$, and define $\Delta t_{max}$ by
Eq. (\ref{dtmax}), where $\Lambda_{min}$ here is the minimum
eigenvalue of the operator ${\cal L}_{QC}$ defined in Eq. (\ref{LQ})
below, then when $\Delta t < \Delta t_{max}$, QCSOM (\ref{QCSOM}) is
guaranteed to converge to the solitary wave $\textbf{u}(\x)$ of Eq.
(\ref{L0UB}) which satisfies the constraints (\ref{Qconstraint}), if
the initial condition is close to $\textbf{u}(\x)$. When $\Delta t >
\Delta t_{max}$, QCSOM (\ref{QCSOM}) diverges.
\end{theorem}

Here $\rm{det}(\hat{\textbf{D}})$ is as defined in Eq. (\ref{Dhat}),
but $Q_j$ is an arbitrary function of $\textbf{u}$ now.

The proof of this theorem is also very similar to that of Theorem
\ref{theorem_g_PC}, thus we will only highlight its differences from
the proof of the latter Theorem. Namely, the principal difference is
that due to the absence of the normalization step in QCSOM
(\ref{QCSOM}), (\ref{muQC}), there appears to be no counterpart of
the orthogonality condition (\ref{orth_g_PC}), which played a
critical role in the proof in Section 4.3. However, as we will see,
a direct counterpart of that condition will arise from different
considerations.

{\bf Proof.} Linearizing the QCSOM, we find that the error satisfies the iteration equation similar to
(\ref{errorun}), except that the iteration operator now becomes
\begin{equation} \label{LQ}
{\cal L}_{QC} \Psi ={\bf M}^{-1}\left( \hat{\bf L}_{PCG} \Psi - h\sum_{j=1}^{r}
\langle \frac{\delta Q_j}{\delta {\bf u}}, \;  \Psi \rangle \frac{\delta Q_j}{\delta {\bf u}} \right),
\end{equation}
where $\hat{\bf L}_{PCG}$ is defined by Eq. (\ref{LPCG}) of Section
4.3. The operator ${\cal L}_{QC}$ can be rewritten as ${\cal
L}_{QC}= {\bf M}^{-1/2} {\cal L}_{QCH} \: {\bf M}^{1/2}$. It is easy
to check that ${\cal L}_{QCH}$ is Hermitian. Using the generalized
Cauchy-Schwartz inequality (\ref{schwartz}), we can also verify that
${\cal L}_{QCH}$ is semi-negative definite. Thus all eigenvalues of
${\cal L}_{QC}$ are real and non-positive, and all eigenfunctions of
${\cal L}_{QC}$ form a complete set. The kernel of ${\cal L}_{QC}$
satisfies the equation
\be
\label{LQCker}
\hat{\bf L}_{PCG}\Psi - h \sum_{j=1}^r \langle \frac{\delta Q_j}{\delta {\bf u}} , \Psi \rangle
 \frac{\delta Q_j}{\delta {\bf u}} =0.
\ee
Notice that {\em both} terms in the above equation are Hermitian and
semi-negative definite operators, thus when taking the inner product
between this equation and $\Psi$, we find that $\Psi$ in the kernel
of ${\cal L}_{QC}$ satisfies Eq. (\ref{L1PsiB}) as well as the
orthogonality relations
\begin{equation} \label{orthB}
\langle \frac{\delta Q_j}{\delta {\bf u}}, \;  \Psi \rangle =0, \hspace{0.5cm} j=1, \dots, r.
\end{equation}
These relations are the counterparts of the orthogonality relations
(\ref{orth_g_PC}) of the previous subsection. Then the proof of this
theorem is completed in exactly the same way as the proof of Theorem
\ref{theorem_g_PC}.

Now we explain the reason for introducing the free parameter $h$
into the scheme (\ref{QCSOM}). Our experience shows that in some
cases (especially when the conserved quantities $C_j$ are large),
the $\delta Q_j/\delta {\bf u}$ terms in Eq. (\ref{LQ}) with $h=1$
cause operator ${\cal L}_{QC}$'s minimum eigenvalue $\Lambda_{min}$
to be large negative --- much larger than that of the operator
$-{\bf M}^{-1} \hat{\bf L}_{QC}$ in magnitude. This forces us to
take very small time steps $\Delta t$ (see Theorem
\ref{theorem_QC}), which severely slows down the convergence speed.
When this happens, our strategy is to introduce a small parameter
$h$ into the scheme (\ref{QCSOM}). The idea is to reduce the effect
of the $\delta Q_j/\delta {\bf u}$ terms on the operator ${\cal
L}_{QC}$, and make its minimum eigenvalue close (or equal) to that
of the operator $-{\bf M}^{-1} \hat{\bf L}_{QC}$. Hence the fast
convergence of the scheme will be restored. This parameter $h$ needs
to be positive so that the relevant terms in Eq. (\ref{LQ}) are
semi-negative definite.

To conclude this section, we point out that the general structures
of the PCSOM (\ref{g_normstep})--(\ref{muQC}) and QCSOM
(\ref{QCSOM}) can also be used to construct the imaginary-time
evolution methods for the case when, in the above notations, $1 < r
< K$. To our knowledge, such a case was not considered in Refs.
\cite{GarciaRipollP_01,Bao_Du,VS_04}, where imaginary-time evolution
methods for vector equations were reported. Of course, the
corresponding imaginary-time evolution methods, unlike the PCSOM
(\ref{g_normstep})--(\ref{muQC}) and QCSOM (\ref{QCSOM}), will {\em
not} be universally-convergent.

\section{Squared-operator iteration methods for isolated solitary waves}

In many dissipative wave systems such as the Ginsburg-Landau
equation, solitary waves exist only when the propagation constants
in the equation take discrete (isolated) values. We call the
corresponding solutions isolated solitary waves. For these isolated
solutions, the propagation constants or powers of the solutions are
unknown and need to be computed together with the solitary wave,
thus the numerical schemes discussed in previous sections do not
apply. In this section, we propose squared-operator iteration
methods for isolated solitary waves.

For simplicity, we consider the case where a vector isolated
solitary wave exists when a single propagation constant takes on
discrete values (other cases can be easily extended). The equation
for the solitary wave can be written as (\ref{L00u}), but the
solution $\textbf{u}(\x)$ now exists only at an unknown discrete
$\mu$ value.
%
We propose the following squared-operator iteration method for these
isolated solitary waves (SOMI):
\begin{equation} \label{SOMI1}
\textbf{u}_{n+1}=\textbf{u}_n-\left[\textbf{M}^{-1}\textbf{L}_1^\dagger
\textbf{M}^{-1}\textbf{L}_0
\textbf{u}\right]_{\textbf{u}=\textbf{u}_n, \; \mu=\mu_n}\Delta t,
\end{equation}
\begin{equation} \label{SOMI2}
\mu_{n+1}=\mu_n+\left. \langle \textbf{u},
\textbf{M}^{-1}\textbf{L}_0 \textbf{u} \rangle
\right|_{\textbf{u}=\textbf{u}_n, \; \mu=\mu_n}\Delta t,
\end{equation}
where $\textbf{L}_1$ is the linear operator as defined in Eq.
(\ref{L1}), and $\textbf{M}$ is a positive-definite and Hermitian
acceleration operator. Linearizing this method around the isolated
solitary wave, we get the iteration equation for the error
$(\tilde{\textbf{u}}_n, \tilde{\mu}_n)$ as
%
%
\begin{equation}
\left( \begin{array}{c} \tilde{\textbf{u}}_{n+1} \\
\tilde{\mu}_{n+1}
\end{array}\right)=(1+\Delta t \: {\cal L}_I) \;
\left( \begin{array}{c} \tilde{\textbf{u}}_n \\ \tilde{\mu}_n
\end{array}\right),
\end{equation}
where
\begin{equation}
{\cal L}_I \; \left( \begin{array}{c} \tilde{\textbf{u}}_n
\\ \tilde{\mu}_n \end{array}\right)
\equiv \left( \begin{array}{c}
-\textbf{M}^{-1}\textbf{L}_1^{\dagger}\textbf{M}^{-1}
\left( \textbf{L}_{1}\tilde{\textbf{u}}_n-\tilde{\mu}_n \textbf{u}\right) \\
\langle \textbf{u}, \;
\textbf{M}^{-1}(\textbf{L}_1\tilde{\textbf{u}}_n-\tilde{\mu}_n
\textbf{u}) \rangle
\end{array}\right).
\end{equation}
It is easy to check that operator ${\cal L}_I$ can be written as
$\mbox{diag}(\textbf{M}^{-1/2}, 1)\:{\cal L}_{IH}\:
\rm{diag}(\textbf{M}^{1/2}, 1)$, where ${\cal L}_{IH}$ is Hermitian
and semi-negative definite. Thus all eigenvalues of ${\cal L}_I$ are
real and non-positive, and all its eigenfunctions form a complete
set. The kernel of ${\cal L}_I$ contains functions $[\Phi(\x), 0]^T$
with $\Phi(\x)$ in the kernel of $\textbf{L}_1$, as well as
functions $[\textbf{F}(\x), 1]^T$ where $\textbf{F}(\x)$ satisfies
the equation
\begin{equation} \label{F}
\textbf{L}_1\textbf{F}=\textbf{u},
\end{equation}
and must be bounded. Assuming that $\textbf{u}$ is not orthogonal to
the kernel of operator $\textbf{L}_1^\dagger$ (which is the generic
case), then Eq. (\ref{F}) has no bounded solution, hence the kernel
of ${\cal L}_I$ only contains the invariance modes $[\Phi({\bf x}),
0]^T$. Then under Assumption 1, SOMI (\ref{SOMI1})-(\ref{SOMI2})
will converge if $\Delta t < -2/\Lambda_{min}$ and diverge if
$\Delta t > -2/\Lambda_{min}$, where $\Lambda_{min}$ is the minimum
eigenvalue of operator ${\cal L}_I$.

Following the motivation for MSOM (\ref{MSOM}) in Sec. 3, we can
construct the modified squared operator method for isolated solitons
in Eq. (\ref{L00u}) as follows (MSOMI):
\begin{equation} \label{MSOMI1}
\textbf{u}_{n+1}=\textbf{u}_n+\left[
-\textbf{M}^{-1}\textbf{L}_1^\dagger \textbf{M}^{-1} \textbf{L}_0
\textbf{u}- \alpha_n \theta_n \textbf{G}_n
\right]_{\textbf{u}=\textbf{u}_n, \; \mu=\mu_n} \Delta t,
\end{equation}
and
\begin{equation} \label{MSOMI2}
\mu_{n+1}=\mu_n+\left[\langle \textbf{u},
\textbf{M}^{-1}\textbf{L}_0 \textbf{u} \rangle - \alpha_n \theta_n
H_n \right]_{\textbf{u}=\textbf{u}_n, \; \mu=\mu_n} \Delta t,
\end{equation}
where
\begin{equation} \label{alphai}
\alpha_n= \frac{1}{\langle \textbf{MG}_n, \textbf{G}_n
\rangle+H_n^2}-\frac{1}{\langle
\left(\textbf{L}_1\textbf{G}_n-H_n\textbf{u}\right),
\; \textbf{M}^{-1}\left(\textbf{L}_1\textbf{G}_n-H_n\textbf{u}\right)\rangle\Delta t},
\end{equation}
\begin{equation} \label{thetai}
\theta_n= -\langle \textbf{L}_1\textbf{G}_n-H_n \textbf{u}, \;
\textbf{M}^{-1} \textbf{L}_0 \textbf{u} \rangle,
\end{equation}
and $(\textbf{G}_n, H_n)$ are functions specified by the user. A
good choice, which is a counterpart of Eq. (\ref{choice2}), is
\begin{equation} \label{choice3}
\textbf{G}_n=\textbf{e}_n \equiv \textbf{u}_n-\textbf{u}_{n-1},
\hspace{0.5cm} H_n=\mu_n-\mu_{n-1}.
\end{equation}
We can show that under similar conditions as in Theorem 2 and the
assumption below Eq. (\ref{F}), this MSOMI also converges for all
isolated solitary waves.


\section{Examples of applications of squared-operator iteration methods}
In this section, we consider various examples of physical interest
to illustrate the performances of the proposed schemes. Most of
these examples are two-dimensional, thus the shooting method can not
work. For some of the examples such as Examples 1 (b,c) and 5, other
numerical methods such as the Petviashvili method and the
imaginary-time evolution method can not work either.

\textbf{Example 6.1 } Consider solitary waves in the two-dimensional NLS equation with a periodic potential,
\begin{equation} \label{NLS2D}
U_{xx}+U_{yy}-V_0 \left(\sin^2x+\sin^2y\right)U + \sigma
|U|^{2}U=-\mu U,
\end{equation}
which arises in nonlinear light propagation in photonic lattices and Bose-Einstein condensation
in optical lattices. Here $\mu$ is the real propagation constant, and $U$ is a complex solution in general.
Writing $U$ into its real and imaginary components, $U(x, y)=u(x, y)+iv(x, y)$,
we get equations for the real functions $u$ and $v$ as
\begin{equation} \label{NLS2Du}
u_{xx}+u_{yy}-V_0 \left(\sin^2x+\sin^2y\right)u + \sigma
(u^2+v^2)u=-\mu u,
\end{equation}
\begin{equation} \label{NLS2Dv}
v_{xx}+v_{yy}-V_0 \left(\sin^2x+\sin^2y\right)v + \sigma
(u^2+v^2)v=-\mu v.
\end{equation}
This system is rotationally invariant, i.e., if $(u, v)^T$ is a solution, so is
\[ \left(\begin{array}{cc} \cos\theta & -\sin\theta \\
\sin\theta & \cos\theta \end{array}\right)\left(\begin{array}{c}u\\v\end{array}\right), \]
where $\theta$ is the angle of rotation (which is constant).
This invariance, in Eq. (\ref{NLS2D}), corresponds to $U \to Ue^{i\theta}$.
This rotational invariance induces an eigenmode $(-v, u)^T$ in the kernel of ${\bf L}_1$.
Clearly, this eigenmode is orthogonal to the solution $(u, v)^T$, satisfying the orthogonality condition in
Theorem \ref{AACITtheorem}.
The kernel of ${\bf L}_1$ does not contain other eigenfunctions (at least in the generic case)
since there are no other invariances in this system.

Solitary waves in Eq. (\ref{NLS2D}) exist only inside the bandgaps
of the system. Thus, the bandgap information is needed first. The
bandgap diagram at various values of $V_0$ is displayed in Fig.
\ref{fig2}(a). For illustration purpose, we fix $V_0=6$, and
determine the solitary waves in different bandgaps below.
\begin{enumerate}
\item[(a)] Vortex solitons in the semi-infinite bandgap under focusing
nonlinearity:

For focusing nonlinearity, $\sigma=1$. In this case, Eq.
(\ref{NLS2D}) admits various types of real and complex solitary-wave
solutions in every bandgap of the system
\cite{YangMuss03,YangStudies,Segev_higherband}. Here we determine a
vortex-soliton solution at $\mu=3$ $(P=14.6004)$, which is marked by
letter 'a' in the semi-infinite bandgap of Fig. 2(a). This solution
is complex valued with a non-trivial phase structure, and it is
displayed in Fig. \ref{fig2}(c, d). Similar solutions have been
reported theoretically in \cite{YangMuss03} before, and have since
been experimentally observed \cite{Chen_vortex,Segev_vortex}. To
determine this solution, we apply the SOM (\ref{SOM}), MSOM
(\ref{MSOM}), (\ref{choice2}), and PCSOM (\ref{PCSOM})-(\ref{mun2})
on Eqs. (\ref{NLS2Du})-(\ref{NLS2Dv}), starting from the initial
condition
\begin{equation} \label{icvortex}
U(x,y)=1.7\left(e^{-x^2-y^2}+e^{-(x-\pi)^2-y^2+i\pi/2}+
e^{-(x-\pi)^2-(y-\pi)^2+i\pi}+e^{-x^2-(y-\pi)^2+3i\pi/2}\right).
\end{equation}
In addition, we choose the acceleration operator ${\bf M}$ as
\begin{equation} \label{bigM}
\textbf{M}=(c-\partial_{xx}-\partial_{yy}) \: \mbox{diag}\: (1, 1).
\end{equation}
The spatial derivatives as well as ${\bf M}^{-1}$ are computed by
the discrete Fourier transform (i.e., by the pseudo-spectral
method). The computational domain is $-6\pi \le x, y \le 6\pi$,
discretized in each dimension by 256 grid points. It should be noted
that the size of the computational domain and the number of grid
points have very little effect on the convergence speed; they mainly
affect the spatial accuracy of the solution. For these three
schemes, we found that the optimal (or nearly optimal) parameters
are $(c, \Delta t)=(3.7, 0.8)$ for SOM and PCSOM, and $(c, \Delta
t)=(3.8, 0.6)$ for MSOM. At these choices of parameters, the error
diagrams versus the number of iterations are displayed in Fig. 2(b).
Here the error is defined as the difference between successive
iteration functions:
\begin{equation}
e_n=\sqrt{\langle U_n-U_{n-1}, U_n-U_{n-1}\rangle}.
\end{equation}
We see that all three schemes converge rather quickly. The convergence speeds of SOM and PCSOM are almost
the same, but MSOM converges much faster. It should be noted that the amount of computations in one
iteration is different for these three methods, with the ratio roughly of $1:1.7:2$ for SOM, PCSOM, and MSOM.
When this factor is also considered, we conclude MSOM converges the fastest, with SOM second, and PCSOM third.

\item[(b)] Solitons in the first bandgap under defocusing
nonlinearity:

Next, we consider solutions in the first bandgap (between the first
and second Bloch bands) under defocusing nonlinearity ($\sigma=-1$).
For this purpose, we pick $\mu=5$, marked by letter 'b' in the first
bandgap of Fig. 2(a). At this point, Eq. (\ref{NLS2D}) admits a
real-valued gap soliton, which is displayed in Fig.
\ref{fig_defoc}(a). Similar solutions have been reported in
\cite{Christodoulides03,KivsharOE} before. To determine this
solution, we apply the SOM (\ref{SOM}), MSOM (\ref{MSOM}),
(\ref{choice2}), and PCSOM (\ref{PCSOM})-(\ref{mun2}) on Eq.
(\ref{NLS2D}), starting from the initial condition
\begin{equation} \label{icdefoc}
U(x,y)=1.15 \: \mbox{sech}(x^2+y^2)\cos(x)\cos(y).
\end{equation}
We take ${\bf M}$ as
\begin{equation} \label{bigM_defoc}
\textbf{M}=c-\partial_{xx}-\partial_{yy}.
\end{equation}
The computational domain is $-5\pi \le x, y \le 5\pi$, discretized
in each dimension by 128 grid points. For these three schemes, we
found that the optimal (or nearly optimal) parameters are $(c,
\Delta t)=(1.8, 0.6)$ for SOM and PCSOM, and $(c, \Delta t)=(2.9,
1.7)$ for MSOM. At these choices of parameters, the error diagrams
versus the number of iterations are displayed in Fig. 3(b). Similar
to the vortex soliton in Fig. 2, we find that the convergence speeds
of SOM and PCSOM are almost the same, but MSOM converges much
faster.

\item[(c)] Vortex solitons in the second bandgap under defocusing
nonlinearity:

We now determine vortex solitons in the second bandgap under
defocusing nonlinearity ($\sigma=-1$). For this purpose, we pick
$\mu=9.4$, marked by letter 'c' in the second bandgap of Fig. 2(a).
At this point, a vortex soliton with distinctive amplitude and phase
distributions exists (see Fig. \ref{fig_array}(a, b)). It is noted
that this vortex soliton is not of the type reported in
\cite{KivsharOE,Segev_higherband}, which lie in the first bandgap
(in our notations). To our knowledge, this type of soliton has never
been reported before in the literature. To determine this solution,
we apply the SOM (\ref{SOM}), MSOM (\ref{MSOM}), (\ref{choice2}),
and PCSOM (\ref{PCSOM})-(\ref{mun2}) on Eqs.
(\ref{NLS2Du})-(\ref{NLS2Dv}). The initial condition is the vortex
solution of these equations at a different value $\mu=9.42$, which
we in turn obtained by the continuation method from a
small-amplitude solution near the edge of the bandgap. The
acceleration operator ${\bf M}$ is the same as (\ref{bigM}). The
computational domain is $-10\pi \le x, y \le 10\pi$, discretized in
each dimension by 256 grid points. For these three schemes, we found
that the optimal (or nearly optimal) parameters are $(c, \Delta
t)=(4.2, 1.7)$ for SOM and PCSOM, and $(c, \Delta t)=(4, 3.1)$ for
MSOM. At these choices of parameters, the error diagrams versus the
number of iterations are displayed in Fig. 4(b). In this case, MSOM
is again the fastest. However, unlike the above two cases, PCSOM
converges much slower than SOM now.
\end{enumerate}

\textbf{Example 6.2} Consider the following system arising from a second-harmonic generation (SHG) model,
\begin{equation} \label{SHG1}
u_{xx}+u_{yy} + uv=\mu u,
\end{equation}
\begin{equation} \label{SHG2}
\frac{1}{2}\left[v_{xx}+5v_{yy}+\frac{1}{2}u^2-v\right]=\mu v.
\end{equation}
Solutions of similar systems have been considered in a number of studies; see, e.g., a recent paper
\cite{LloydC04} and references therein. Note that in the original reduction from the SHG model, the
right hand side of the $v$-equation is usually $2\mu v$. But in order to cast the equations into the
form (\ref{L00u}),
we have divided the $v$-equation by 2, so that we work with Eq. (\ref{SHG2}) instead.
Here we deliberately make the
$v_{xx}$ and $v_{yy}$ coefficients different, so that the radial symmetry is broken, hence this system
is not reducible to a radially symmetric (and hence essentially one-dimensional) problem.
At $\mu=0.1$, this system admits a solitary wave with total power
$P=\langle u, u\rangle+\langle v, v\rangle=47.3744$, which is displayed in Fig. 5(a, b).
We take the initial condition as
\begin{equation}
u(x, y)=\mbox{sech}(0.35\sqrt{x^2+y^2}), \hspace{0.5cm} v=0.3\mbox{sech}(0.5\sqrt{x^2+y^2}).
\end{equation}
The acceleration operator ${\bf M}$ is taken as
\begin{equation}
\label{M_Ex62}
{\bf M}=\mbox{diag}\left[\mu-\partial_{xx}-\partial_{yy}, \; \mu+\frac{1}{2}-
\frac{1}{2} \partial_{xx}-\frac{5}{2} \partial_{yy}\right],
\end{equation}
where the choice of constants $\mu$ and $\mu+\frac12$ is motivated
by our earlier results on optimal accelerations for imaginary-time
evolution methods \cite{YangLakoba}. The computational domain is
$-25 \le x, y, \le 25$, and the number of grid points along each
dimension is 64. The iteration results of SOM (\ref{SOM}), MSOM
(\ref{MSOM}), (\ref{choice2}) and PCSOM (\ref{PCSOM})-(\ref{mun2})
at optimal (or nearly optimal) $\Delta t$ values 0.37, 0.59, and
0.63 respectively are displayed in Fig. 5(c). Again, MSOM delivers
the best performance, and PCSOM is the slowest.

\textbf{Example 6.3} The next example is the coupled two-dimensional NLS equations with saturable nonlinearity,
\begin{equation} \label{CPNLS1}
u_{xx}+u_{yy}+\frac{u^2+v^2}{1+s(u^2+v^2)}u=\mu_1 u,
\end{equation}
\begin{equation} \label{CPNLS2}
v_{xx}+v_{yy}+\frac{u^2+v^2}{1+s(u^2+v^2)}v=\mu_2 v,
\end{equation}
which has been studied, e.g., in \cite{GarciaRipollP_01, Kivshar_dipole, Yang_dipole}.
Here $s$ is the saturation constant. At $s=0.5, \mu=1, \mu_2=0.5$
($P_1=85.3884, P_2=29.1751$), this system admits a solution whose $u$-component is single-humped,
but its $v$-component is a dipole state. This solution is displayed in Fig. 6(a, b).
Note that this solution is not radially symmetric, thus is not reducible to a one-dimensional problem.
Taking the initial condition as
\begin{equation}
u(x, y)=3e^{-0.2r^2}, \hspace{0.5cm} v(x, y)=1.5 r \: e^{-0.2r^2}\cos\theta,
\end{equation}
where $(r, \theta)$ are the polar coordinates,
the acceleration operator ${\bf M}$ as
\begin{equation}
\label{M_Ex63}
{\bf M}=\mbox{diag}\left[\mu_1-\partial_{xx}-\partial_{yy}, \; \mu_2-\partial_{xx}-\partial_{yy}\right],
\end{equation}
the computational domain as $-12 \le x, y, \le 12$, and the number
of grid points along the two dimensions as 64, the iteration results
of SOM (\ref{SOM}), MSOM (\ref{MSOM}), (\ref{choice2}) and PCSOM
(\ref{PCSOM_vector})-(\ref{PCSOM3_vector}) at optimal $\Delta t$
values 1.9, 2.65, and 1.85 are displayed in Fig. 6(c). As in
(\ref{M_Ex62}) above, the choice of the constants $\mu_1$ and
$\mu_2$ in the acceleration operator (\ref{M_Ex63}) is also
motivated by our previous studies on the accelerated imaginary-time
evolution method \cite{YangLakoba}. Again, MSOM delivers the best
performance, and PCSOM is the slowest.

\textbf{Example 6.4} The next example is intended to compare the
performances of PCSOM (\ref{g_normstep})--(\ref{muQC}) and QCSOM
(\ref{QCSOM}). This example comes from the three-wave interaction
system \cite{3w1,3w2} and has the form
\begin{equation} \label{threewave1}
u_{xx}+u_{yy} + vw=\mu_1 u,
\end{equation}
\begin{equation}  \label{threewave2}
v_{xx}+v_{yy} + uw=\mu_2 v,
\end{equation}
\begin{equation}  \label{threewave3}
w_{xx}+w_{yy} + uv=(\mu_1+\mu_2)w,
\end{equation}
where $u, v$ and $w$ are real functions, and $\mu_1$ and $\mu_2$ are propagation constants.
At $\mu_1=0.5$ and $\mu_2=1$, this system has a radially symmetric solution displayed in Fig.
\ref{threewavefig}(a), and
\begin{equation}
Q_1\equiv \langle u, u\rangle+\langle w, w\rangle=66.3096, \hspace{0.5cm}
Q_2\equiv \langle v, v\rangle+\langle w, w\rangle=47.2667.
\end{equation}
Here we want to determine this solution with the pre-specified
quantities $Q_1$ and $Q_2$ as above. Note that this problem is of
the form (\ref{L0U_quadr})-(\ref{Qconstraint_quadr}) and
(\ref{L0UB})-(\ref{Qconstraint}), but not of the form (\ref{L00u})
or (\ref{L0U_vector}). Taking the initial condition as
\begin{equation}
u(x, y)=2.5\mbox{sech}0.8r, \hspace{0.5cm} v(x, y)=2.2\mbox{sech}0.8r, \hspace{0.5cm}
w(x, y)=1.9\mbox{sech}0.8r,
\end{equation}
where $r=\sqrt{x^2+y^2}$, the acceleration operator ${\bf M}$ as
\begin{equation}
{\bf M}=\mbox{diag}\left[\mu_1-\partial_{xx}-\partial_{yy}, \; \mu_2-\partial_{xx}-\partial_{yy}, \;
\mu_1+\mu_2-\partial_{xx}-\partial_{yy}  \right],
\end{equation}
the computational domain as $-15 \le x, y, \le 15$, the number of
grid points along the two dimensions as 64, and the parameter $h$ in
the QCSOM as $h=0.01$, the iteration results of PCSOM
(\ref{g_normstep})--(\ref{muQC}) and QCSOM (\ref{QCSOM}) at the
(same) optimal value $\Delta t=0.49$ are displayed in Fig.
\ref{threewavefig}(b). This figure shows that the QCSOM converges
slightly slower than the PCSOM. However, it is noted that each QCSOM
iteration involves less computations than the PCSOM, thus we
conclude that the performances of PCSOM and QCSOM are comparable.

\textbf{Example 6.5} The Ginzburg-Landau equation is of the form
\begin{equation} \label{GL0}
i\Phi_t+(1-\gamma_1 i) \Phi_{xx}-i\gamma_0 \Phi+|\Phi|^2\Phi=0,
\end{equation}
where $\gamma_0$ is the damping/pumping coefficient (when $\gamma_0$
is negative/positive). We seek solitary waves in this equation of
the form
\begin{equation}
\Phi(x, t)=U(x)e^{i\mu t},
\end{equation}
where $\mu$ is a real propagation constant, then function $U(x)$ satisfies the equation
\begin{equation} \label{GL}
(1-\gamma_1 i) U_{xx}-i\gamma_0 U+|U|^2U=\mu U.
\end{equation}
If $\gamma_0$ or $\gamma_1$ is non-zero, then solitary waves in this
equation are always complex-valued, and they can only exist at
isolated propagation constants. When $\gamma_0=0.3$ and
$\gamma_1=1$, the solitary wave, which exists at the discrete value
$\mu=1.2369$, is plotted in Fig. \ref{fig_GL}(a). Writing this
equation into two real-valued equations for [Re($U$), Im($U$)], and
applying SOMI (\ref{SOMI1})-(\ref{SOMI2}) or MSOMI
(\ref{MSOMI1})-(\ref{choice3}) with $M=c-\partial_{xx}$ and initial
conditions $u_0(x)=1.6\mbox{sech}(x)$, $\mu_0=1.2$, we can obtain
this isolated solitary wave. At optimal scheme parameters $(c,
\Delta t)=(1.6, 0.3)$ for SOMI and $(c, \Delta t)=(1.4, 0.12)$ for
MSOMI, the error diagram is displayed in Fig. \ref{fig_GL}(b). Here
the error is defined as
\[e_n=\sqrt{\langle u_n-u_{n-1},
u_n-u_{n-1}\rangle}+|\mu_n-\mu_{n-1}|.\] We see that MSOMI converges
much faster than SOMI, which is consistent with previous numerical
experiments.

We should point out that in this example, since $\gamma_0>0$, the
soliton we obtained is unstable (because the zero background is
unstable). So this soliton can not be obtained by simulating the
Ginzburg-Landau equation (\ref{GL0}). However, our proposed method
SOMI/MSOMI can produce this unstable solution quite easily.



\section{Summary}

In this paper, we have developed three iteration methods
--- the squared operator method (SOM), the modified squared operator method (MSOM),
and the power(or any quantity)-conserving squared operator method
(PCSOM/QCSOM), for computing solitary waves in general nonlinear
wave equations. The solitary waves can exist at either continuous or
discrete propagation constants. These numerical methods are based on
iterating new differential equations whose linearization operators
are squares of those for the original equations. We proved that all
these methods are guaranteed to converge to all types of solitary
waves as long as the time step in the iteration schemes is below a
certain threshold value. Due to the use of acceleration techniques,
these methods are fast converging. Since these methods are
compatible with the pseudo-spectral method, their spatial accuracy
is exponentially high. Furthermore, these methods can treat problems
in arbitrary dimensions with little change in the programming, and
they are very easy to implement. To test the relative performances
of these methods, we have applied them to various solitary wave
problems of physical interest, such as higher-gap vortex solitons in
the two-dimensional nonlinear Schr\"odinger equations with periodic
potentials and isolated solitons in Ginzburg-Landau equations. We
found that MSOM delivers the best performance among all the methods
proposed.

Even though MSOM delivers the best performance, SOM and PCSOM/QCSOM
have their own advantages as well. For instance, SOM is simpler to
implement. PCSOM/QCSOM would be advantageous if the problem at hand
specifies the power or other conserved quantity of the solution
rather than the propagation constants. In addition, PCSOM/QCSOM
works for linear eigenvalue problems (by setting the linear
eigenfunction to have a fixed norm), while SOM and MSOM do not. In
some cases, SOM or PCSOM is more tolerant to the choice of initial
conditions, i.e., they converge for a larger range of initial
conditions than MSOM. Thus all the methods developed in this paper
can be useful for different purposes, and the reader can pick them
judiciously depending on the problem at hand.

It is noted that these methods can deliver good performance even
with suboptimal choices of scheme parameters. But how to get the
best performance out of these schemes (i.e., how to find optimal
scheme parameters) is still an important open question. For the
Petviashvili method, conditions for optimal or nearly optimal
convergence have been studied in \cite{Peli_Pet,LY06}. For the
accelerated imaginary-time evolution method, optimal acceleration
parameters have been obtained for a large class of equations
\cite{YangLakoba}.  Such results can help us to select the scheme
parameters for the squared-operator methods in this paper (as we
have done in Examples 6.2--6.4). But a rigorous and comprehensive
study on optimal scheme parameters for the squared-operator methods
proposed in this paper is a non-trivial issue and will be left for
future studies.


\section*{Acknowledgment}
Part of this work was done when J.Y. was visiting Zhou Pei-Yuan Center
for Applied Mathematics at Tsinghua University. The hospitality of
the Center was appreciated. The work of J.Y. was supported in part
by AFOSR under grant USAF 9550-05-1-0379, and the work of T.I.L. was supported in
part by NSF under grant DMS-0507429.

\def\theequation {A.\arabic{equation}}
\begin{center}
{\bf Appendix:  Families of squared-operator methods}
\end{center}
The squared operator methods proposed in this paper were based on
the operator ${\cal L}$ given in Eq. (\ref{curlL}). It turns out
that one can construct a family of squared operator methods which
contain the methods in this paper as particular cases. Consider the
following squared operator iteration method for Eq. (\ref{L0U}):
\begin{equation} \label{SOMF}
\textbf{u}_{n+1}=\textbf{u}_n-\left[\textbf{M}^{-a}\textbf{L}_1^\dagger
\textbf{M}^{-b} \textbf{L}_0
\textbf{u}\right]_{\textbf{u}=\textbf{u}_n}\Delta t,
\end{equation}
where $\textbf{M}$ is a positive definite Hermitian operator, and
$a$ and $b$ are arbitrary constants. The linearized equation of this
method for the error $\tilde{\textbf{u}}_n$ is still Eq.
(\ref{error0}), except that ${\cal L}$ is replaced by
\begin{equation}
\label{calL_f}
{\cal L}_f \equiv
-\textbf{M}^{-a}\textbf{L}_1^\dagger\textbf{M}^{-b} \textbf{L}_1
\end{equation}
now. This operator can be rewritten as
\begin{equation}
{\cal L}_f =
-\textbf{M}^{-a/2}\left(\textbf{M}^{-b/2}\textbf{L}_1\textbf{M}^{-a/2}\right)^\dagger
\left(\textbf{M}^{-b/2}\textbf{L}_1\textbf{M}^{-a/2}\right)\textbf{M}^{a/2},
\end{equation}
thus its eigenvalues are clearly all non-positive. Repeating the
proof in Sec. \ref{sec_SO}, we can readily show that the SOM
(\ref{SOMF}) is guaranteed to converge if $\Delta t <
-2/\Lambda_{min}$, where $\Lambda_{min}$ is the minimum eigenvalue
of operator ${\cal L}_f$. If we choose $a=1$ and $b=1$, then method
(\ref{SOMF}) becomes SOM (\ref{SOM}). For the family of SOMs
(\ref{SOMF}), the corresponding MSOMs, PCSOMs, as well as methods
for isolated solitary waves can be readily constructed. Below we
illustrate such a construction for a particular choice of $a$ and
$b$. Namely, we consider the member of family (\ref{SOMF}) with
$a=0$ and $b=2$, whose implementation of each iteration requires
slightly fewer operations than the implementation of method
(\ref{SOM}) in the main text. In this case, the corresponding
squared operator methods are listed below.
\begin{itemize}
\item SOM for Eq. (\ref{L0U}):
\begin{equation} \label{SOM2}
\textbf{u}_{n+1}=\textbf{u}_n-\left[\textbf{L}_1^\dagger
\textbf{M}^{-2} \textbf{L}_0
\textbf{u}\right]_{\textbf{u}=\textbf{u}_n}\Delta t.
\end{equation}
\item MSOM for Eq. (\ref{L0U}):
\begin{equation} \label{MSOM2}
\textbf{u}_{n+1}=\textbf{u}_n-\left[\textbf{L}_1^\dagger
\textbf{M}^{-2} \textbf{L}_0 \textbf{u} - \alpha_n \langle \textbf{G}_n, \;
\textbf{L}_1^\dagger \textbf{M}^{-2} \textbf{L}_0 \textbf{u}
\rangle \textbf{G}_n
\right]_{\textbf{u}=\textbf{u}_n}\Delta t,
\end{equation}
where
\begin{equation} \label{alpha2}
\alpha_n= \frac{1}{\langle \textbf{G}_n,\textbf{G}_n
\rangle}-\frac{1}{\langle \textbf{M}^{-1}\textbf{L}_1\textbf{G}_n,
\; \textbf{M}^{-1}\textbf{L}_1\textbf{G}_n\rangle\Delta t},
\end{equation}
and $\textbf{G}_n$ is a user-specified function such as
(\ref{choice2}).
\item PCSOM for Eq. (\ref{L00u}):
\begin{equation}
\label{A7}
{\bf u}_{n+1}=\left[\frac{P}{\langle \hat{\bf u}_{n+1}, \hat{\bf u}_{n+1}
\rangle}\right]^{\frac{1}{2}} \hat{\bf u}_{n+1},
\end{equation}
\begin{equation} \label{A8}
\hat{\bf u}_{n+1}={\bf u}_n- \left[{\bf L}_1^\dagger {\bf M}^{-2}{\bf L}_0{\bf u}-\gamma
{\bf u}\right]_{{\bf u}={\bf u}_n, \: \mu=\mu_n} \Delta t,
\end{equation}
and
\begin{equation}
\gamma=\frac{\langle {\bf u}, \; {\bf L}_1^\dagger {\bf M}^{-2}{\bf L}_0{\bf u} \rangle
}{\langle {\bf u}, \;  {\bf u} \rangle }, \hspace{0.5cm}
\mu_n=\left. \frac{\langle {\bf u}, \; {\bf M}^{-2}{\bf L}_{00}{\bf u} \rangle}{\langle {\bf u}, \;
{\bf M}^{-2}{\bf u} \rangle }\right|_{{\bf u}={\bf u}_n}.
\end{equation}
We note that for this power-conserving scheme, the $\gamma$ term in
Eq. (\ref{A8}) can be dropped due to the presence of the power
normalization step (\ref{A7}). It is easy to check that the reduced
scheme has the same linearized iteration operator as the original
one above, thus possesses the same convergence properties. However,
this can not be done for the PCSOM (\ref{PCSOM})-(\ref{mun2}) in the
main text if ${\bf M}\ne 1$. The other forms of the PCSOM for the
cases (\ref{L0U_vector}) and
(\ref{L0U_quadr})-(\ref{Qconstraint_quadr}) can be similarly written
down.
\item QCSOM for Eqs. (\ref{L0UB})-(\ref{Qconstraint}):
\begin{equation} \label{QCSOMb}
{\bf u}_{n+1}={\bf u}_n- \left[{\bf L}_1^\dagger {\bf M}^{-2}{\bf
L}_0{\bf u} + h\: \sum_{j=1}^{r} \left(Q_j({\bf
u})-C_j\right)\frac{\delta Q_j}{\delta {\bf u}} \right]_{{\bf
u}={\bf u}_n, \: \vec{\mu}=\vec{\mu}_n} \Delta t,
\end{equation}
where
\begin{equation} \label{muQCb}
\vec{\mu}_n = \left. \langle {\bf B}, \; {\bf M}^{-2}{\bf B}
\rangle^{-1} \langle {\bf B}, {\bf M}^{-2}{\bf L}_{00}{\bf u}
\rangle\right|_{{\bf u}={\bf u}_n}.
\end{equation}
%
\item SOMI for isolated solitary waves in Eq. (\ref{L00u}):
\begin{equation}
\textbf{u}_{n+1}=\textbf{u}_n-\left[\textbf{L}_1^\dagger
\textbf{M}^{-2}\textbf{L}_0
\textbf{u}\right]_{\textbf{u}=\textbf{u}_n, \; \mu=\mu_n}\Delta t,
\end{equation}
\begin{equation}
\mu_{n+1}=\mu_n+\left. \langle \textbf{u},
\textbf{M}^{-2}\textbf{L}_0 \textbf{u} \rangle
\right|_{\textbf{u}=\textbf{u}_n, \; \mu=\mu_n}\Delta t,
\end{equation}
\item MSOMI for isolated solitary waves in Eq. (\ref{L00u}):
\begin{equation}
\textbf{u}_{n+1}=\textbf{u}_n+\left[-\textbf{L}_1^\dagger
\textbf{M}^{-2} \textbf{L}_0 \textbf{u}- \alpha_n \theta_n
\textbf{G}_n \right]_{\textbf{u}=\textbf{u}_n, \; \mu=\mu_n} \Delta
t,
\end{equation}
\begin{equation}
\mu_{n+1}=\mu_n+\left[\langle \textbf{u},
\textbf{M}^{-2}\textbf{L}_0 \textbf{u} \rangle - \alpha_n \theta_n
H_n \right]_{\textbf{u}=\textbf{u}_n, \; \mu=\mu_n} \Delta t,
\end{equation}
where
\begin{equation}
\alpha_n= \frac{1}{\langle \textbf{G}_n,\textbf{G}_n
\rangle+H_n^2}-\frac{1}{\langle
\textbf{M}^{-1}\left(\textbf{L}_1\textbf{G}_n-H_n\textbf{u}\right),
\;
\textbf{M}^{-1}\left(\textbf{L}_1\textbf{G}_n-H_n\textbf{u}\right)\rangle\Delta
t},
\end{equation}
\begin{equation}
\theta_n= -\langle \textbf{G}_n, \; \textbf{L}_1^\dagger
\textbf{M}^{-2} \textbf{L}_0 \textbf{u} \rangle+\langle
H_n\textbf{u}, \textbf{M}^{-2}\textbf{L}_0 \textbf{u} \rangle,
\end{equation}
and $(\textbf{G}_n, H_n)$ are user specified functions such as
(\ref{choice3}).
\end{itemize}
All these methods can be shown to have similar convergence
properties as their counterparts in the main text.

As we have already noted, in all the methods presented in this
Appendix starting with Eq. (\ref{SOM2}), each iteration involves a
little less computations than their counterparts in the methods
presented in the main text. However, we did not advocate for these
methods for two reasons. First, we can show that the convergence of
these methods is always slower than that of their counterparts in
the text when $\textbf{L}_1$ is Hermitian. In such a case, using
Theorem 5.6.9 in \cite{HornJohnson91}, we can show that
$\Lambda_{min}/\Lambda_{max}$ is smaller for ${\cal L}$ than it is
for ${\cal L}_f$, hence SOM (\ref{SOM}) converges faster than its
counterpart (\ref{SOM2}) in view of Eq. (\ref{R*}). Second, for the
squared-operator methods in the main text, we can carry out explicit
convergence analysis on some familiar examples such as the NLS
equation. This would not be possible for the methods considered in
this Appendix. Overall, our numerical testing shows that the
squared-operator methods presented in the main text give the best
performance among the family of methods (\ref{SOMF}) with other
choices of $a$ and $b$.

\newpage

\begin{figure}[h]
\begin{center}
\parbox{10cm}{\postscript{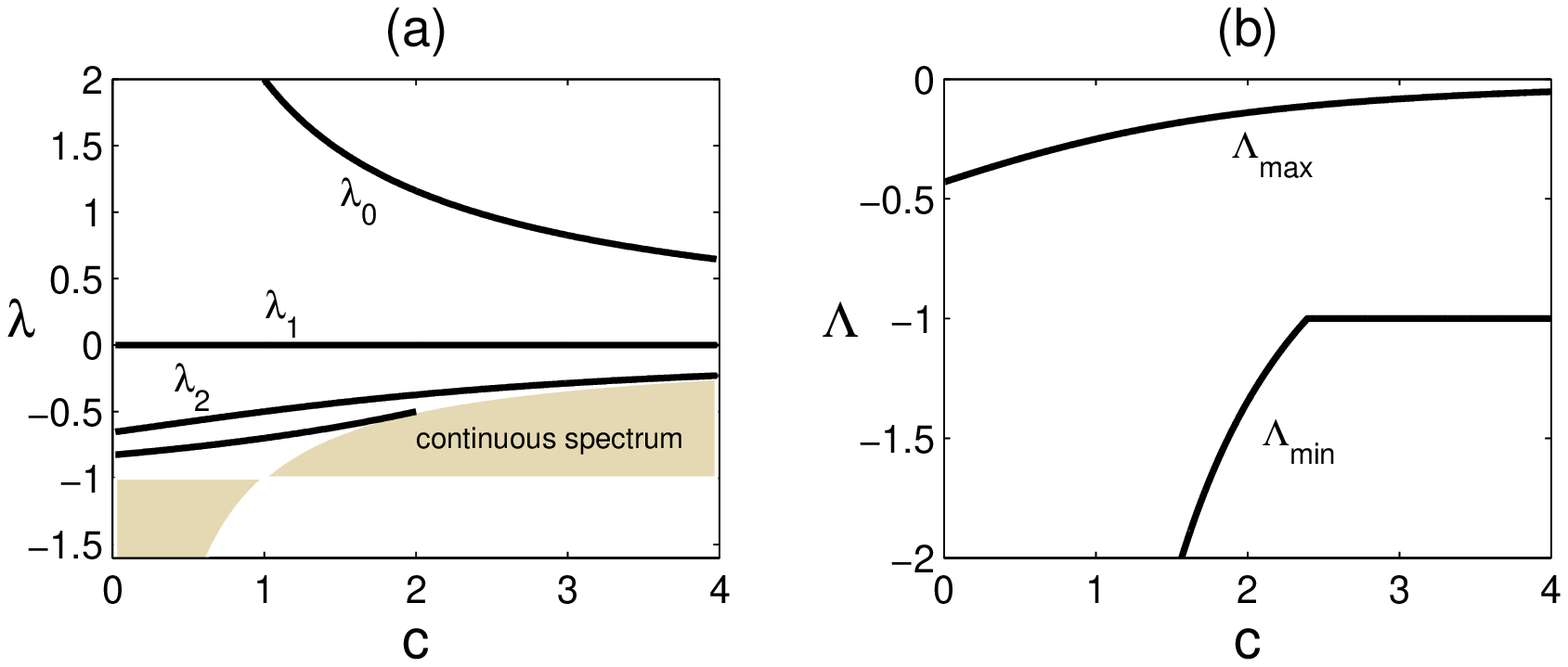}{1.0}}

\parbox{10cm}{\postscript{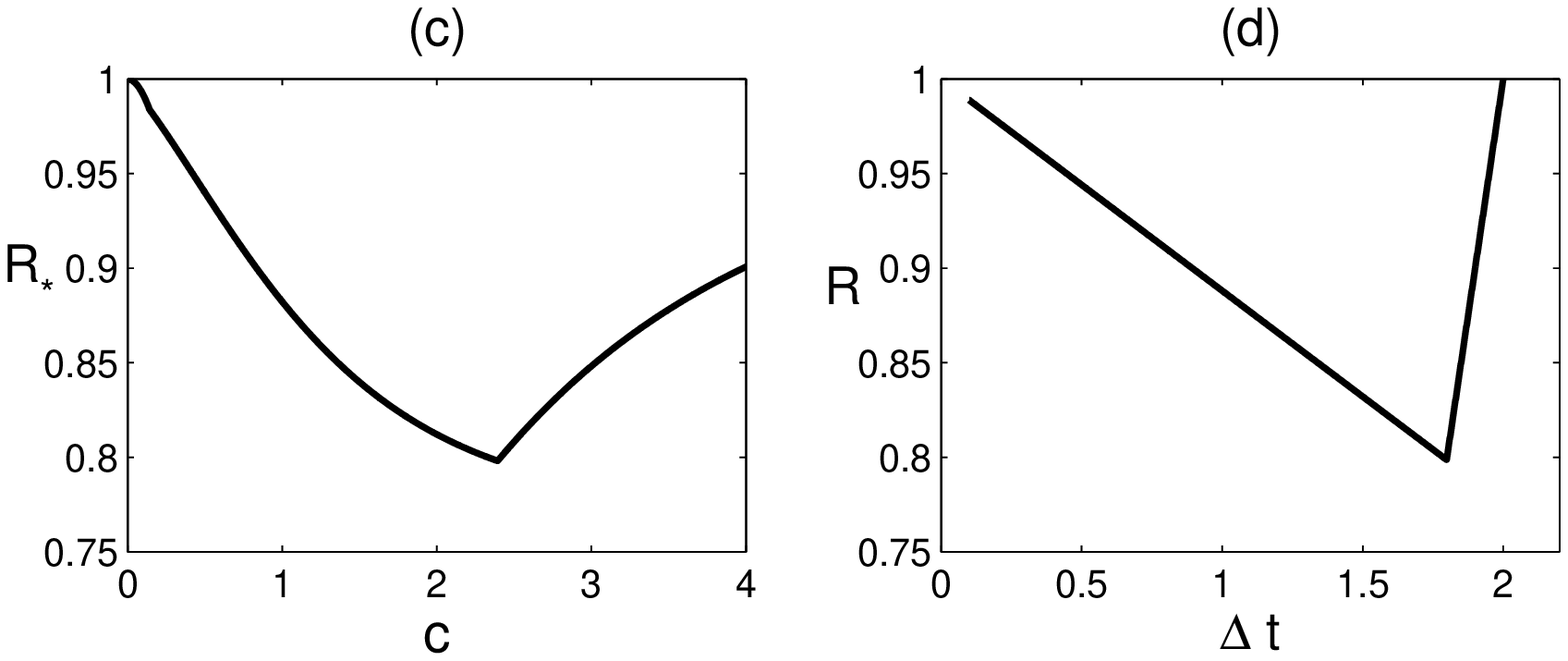}{1.0}}
\caption{Analysis of the SOM (\ref{SOM}) applied to the soliton
(\ref{soliton}) of the NLS equation (\ref{NLS}). (a) Eigenvalues of
operator $M^{-1}L_1$, with $M$ given in (\ref{M_form}), versus the
acceleration parameter $c$; (b) maximum and minimum (nonzero)
eigenvalues of the iteration operator ${\cal
L}=-\left(M^{-1}L_1\right)^2$; (c) graph of convergence factor
$R_*(c)$ versus $c$; its minimum occurs at $c=c_{opt}=6-\sqrt{13}$;
(d) convergence factor function $R(\Delta t; c)$ versus $\Delta t$
at $c=c_{opt}$. \label{fig1}}
\end{center}
\end{figure}

\begin{figure}[h]
\begin{center}
\parbox{12cm}{\postscript{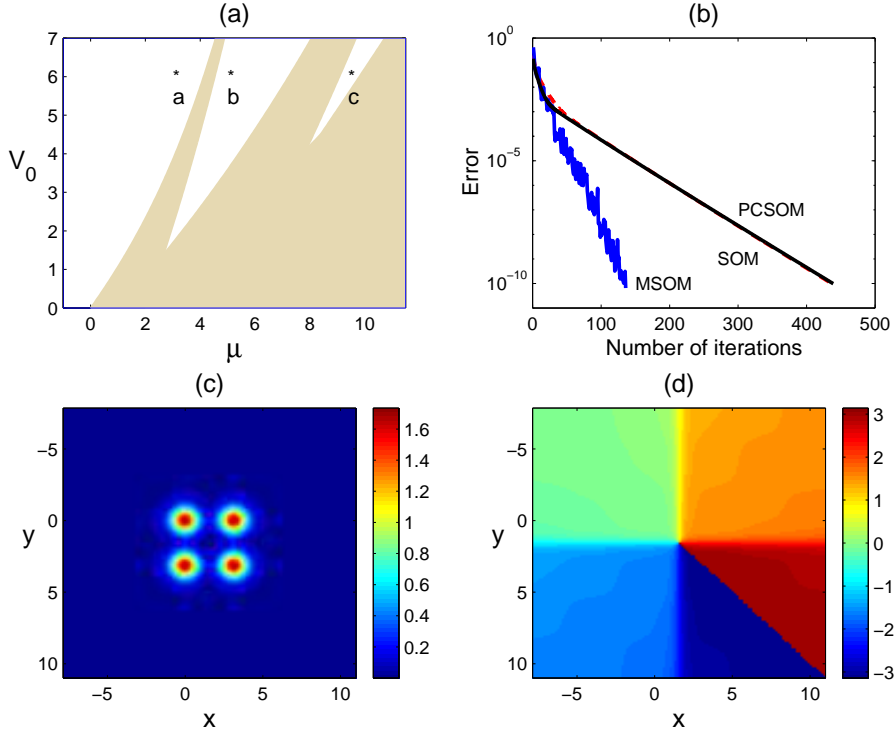}{1.0}}

\caption{ (a) Bandgap structure of Eq. (\ref{NLS2D}); (b) error
diagrams for SOM (\ref{SOM}), MSOM (\ref{MSOM}), (\ref{choice2}) and
PCSOM (\ref{PCSOM})-(\ref{mun2}) at optimal $c$ and $\Delta t$
values (see text); SOM and PCSOM are almost indistinguishable; (c,
d) a vortex soliton in the semi-infinite bandgap of Eq.
(\ref{NLS2D}) with focusing nonlinearity. Here $\mu=3$
($P=14.6004$), marked by letter 'a' in panel (a). (c) is the
amplitude ($|U|$) distribution, and (d) the phase distribution.
\label{fig2}}
\end{center}
\end{figure}

\begin{figure}[h]
\begin{center}
\parbox{12cm}{\postscript{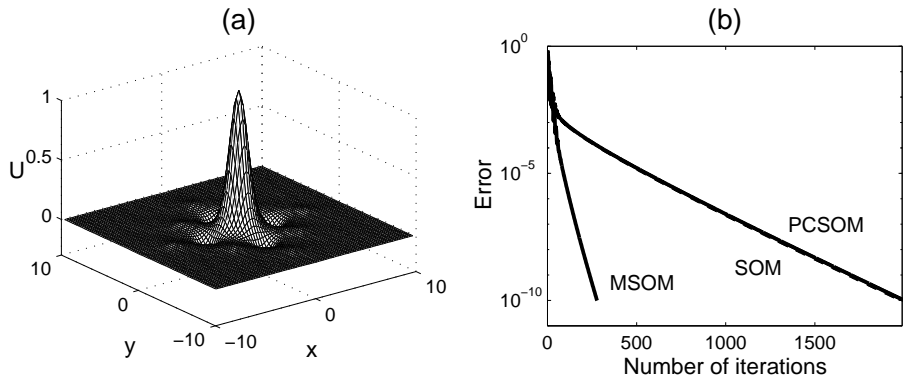}{1.0}}

\caption{(a) A solitary wave in the first bandgap of Eq.
(\ref{NLS2D}) with defocusing nonlinearity at $\mu=5$ ($P=2.4936$),
marked by letter 'b' in Fig. \ref{fig2}(a); (b) error diagrams for
SOM (\ref{SOM}), MSOM (\ref{MSOM}), (\ref{choice2}) and PCSOM
(\ref{PCSOM})-(\ref{mun2}) at optimal $c$ and $\Delta t$ values (see
text); SOM and PCSOM are almost indistinguishable. \label{fig_defoc}
}
\end{center}
\end{figure}

\begin{figure}[h]
\begin{center}
\parbox{12cm}{\postscript{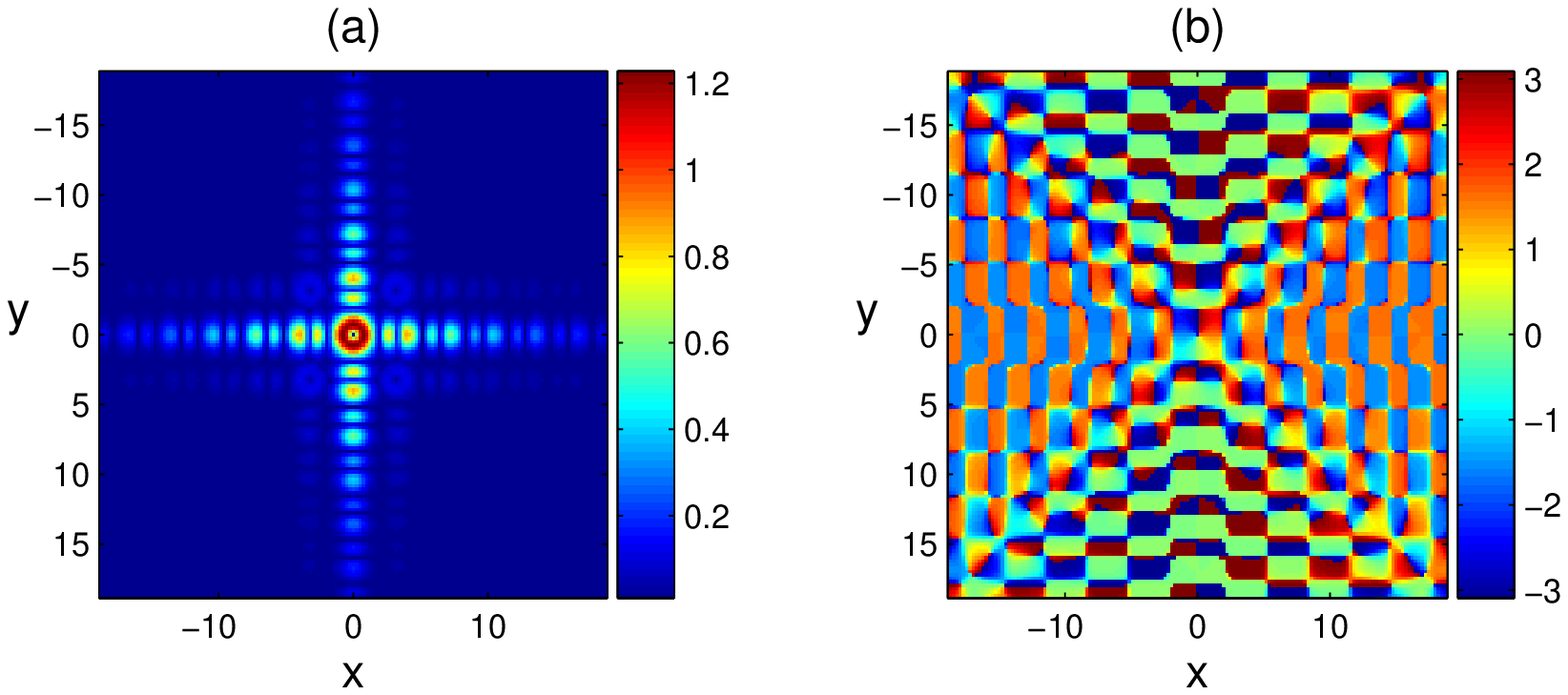}{1.0}}

\parbox{6cm}{\postscript{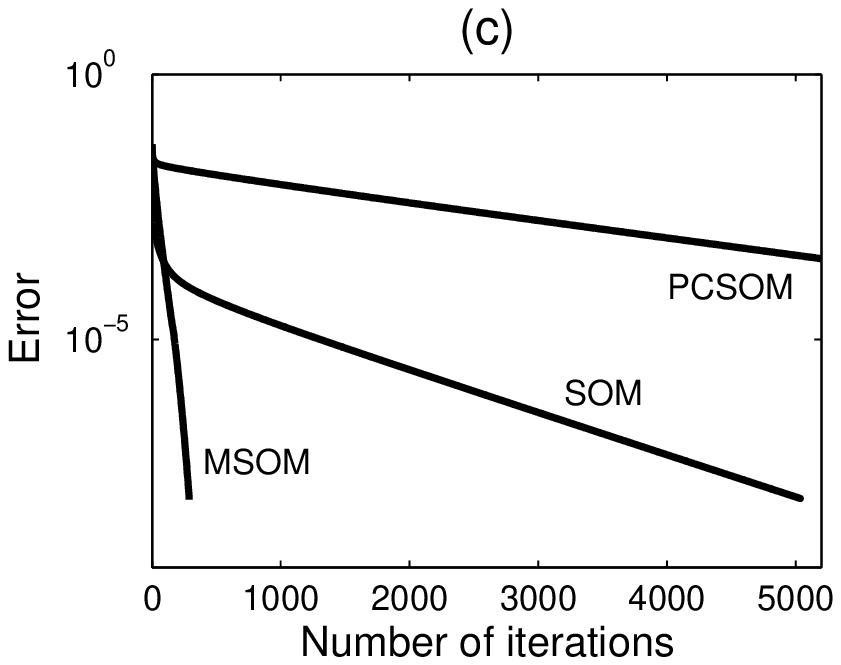}{1.0}}
\caption{  (a, b) A vortex-soliton solution in the second bandgap of
Eq. (\ref{NLS2D}) with defocusing nonlinearity at $\mu=9.4$
$(P=18.3578)$, marked by letter 'c' in Fig. \ref{fig2}(a). (a):
amplitude distribution; (b) phase distribution. (c) error diagrams
for SOM (\ref{SOM}), MSOM (\ref{MSOM}), (\ref{choice2}) and PCSOM
(\ref{PCSOM})-(\ref{mun2}) at optimal $c$ and $\Delta t$ values (see
text). \label{fig_array} }
\end{center}
\end{figure}

\begin{figure}[h]
\begin{center}
\parbox{12cm}{\postscript{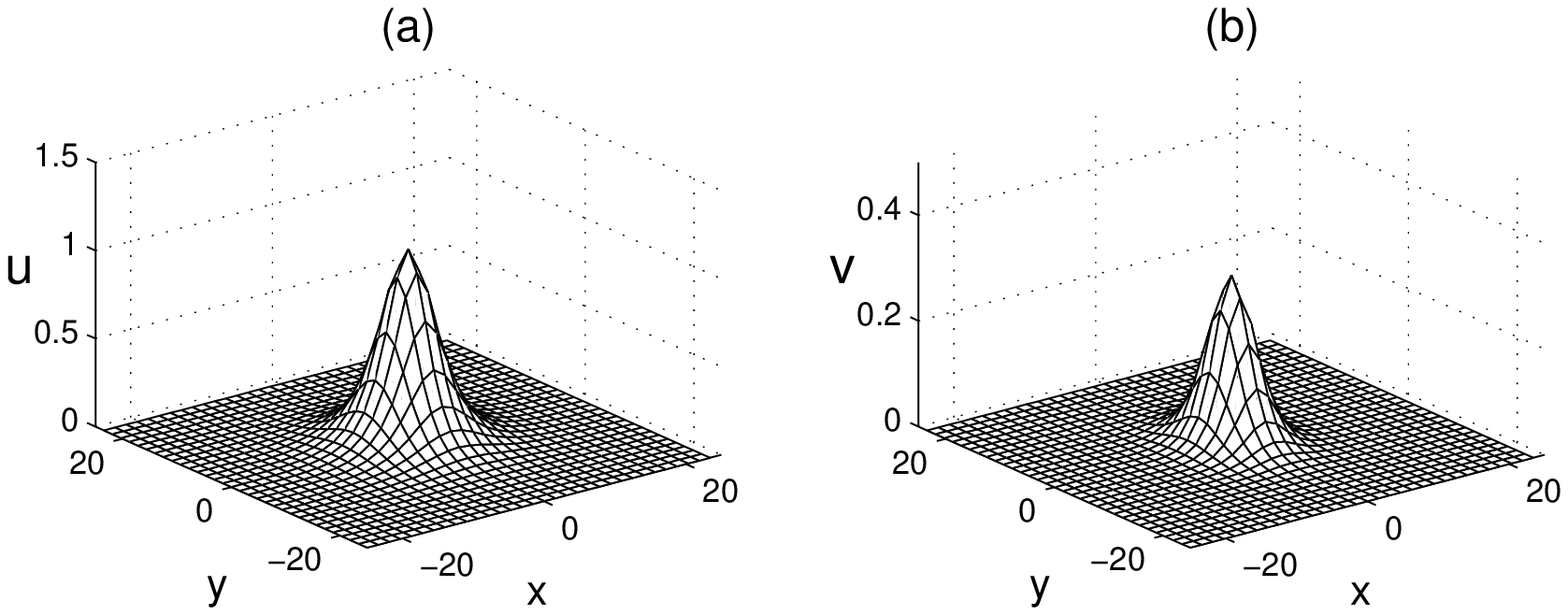}{1.0}}

\parbox{6cm}{\postscript{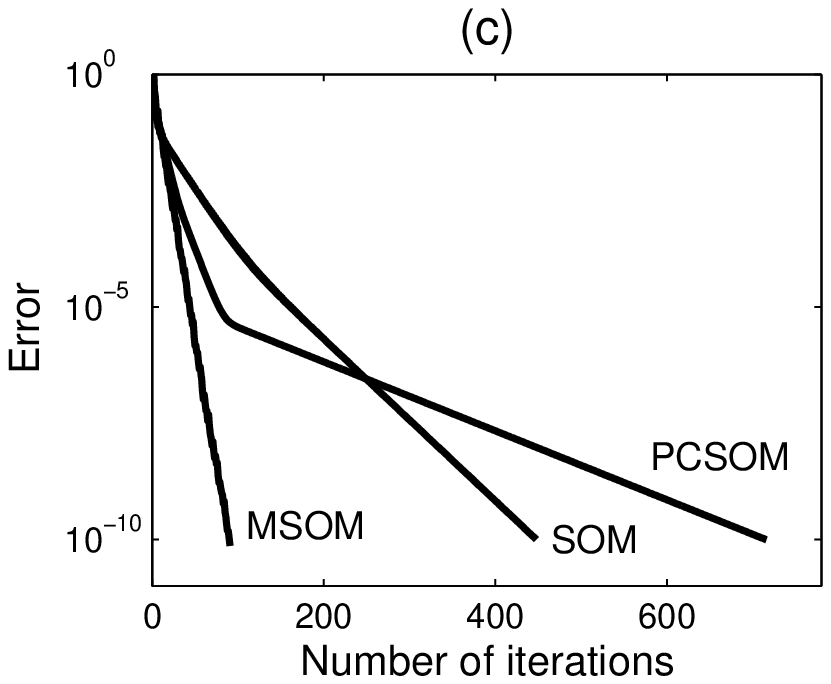}{1.0}}
\caption{(a, b) A fundamental solitary wave in the SHG model
(\ref{SHG1})-(\ref{SHG2}) at $\mu=0.1$ ($P=47.3744$); (c) error
diagrams for SOM (\ref{SOM}), MSOM (\ref{MSOM}), (\ref{choice2}) and
PCSOM (\ref{PCSOM})-(\ref{mun2}) at optimal scheme parameters (see
text).    }
\end{center}
\end{figure}

\begin{figure}[h]
\begin{center}
\parbox{12cm}{\postscript{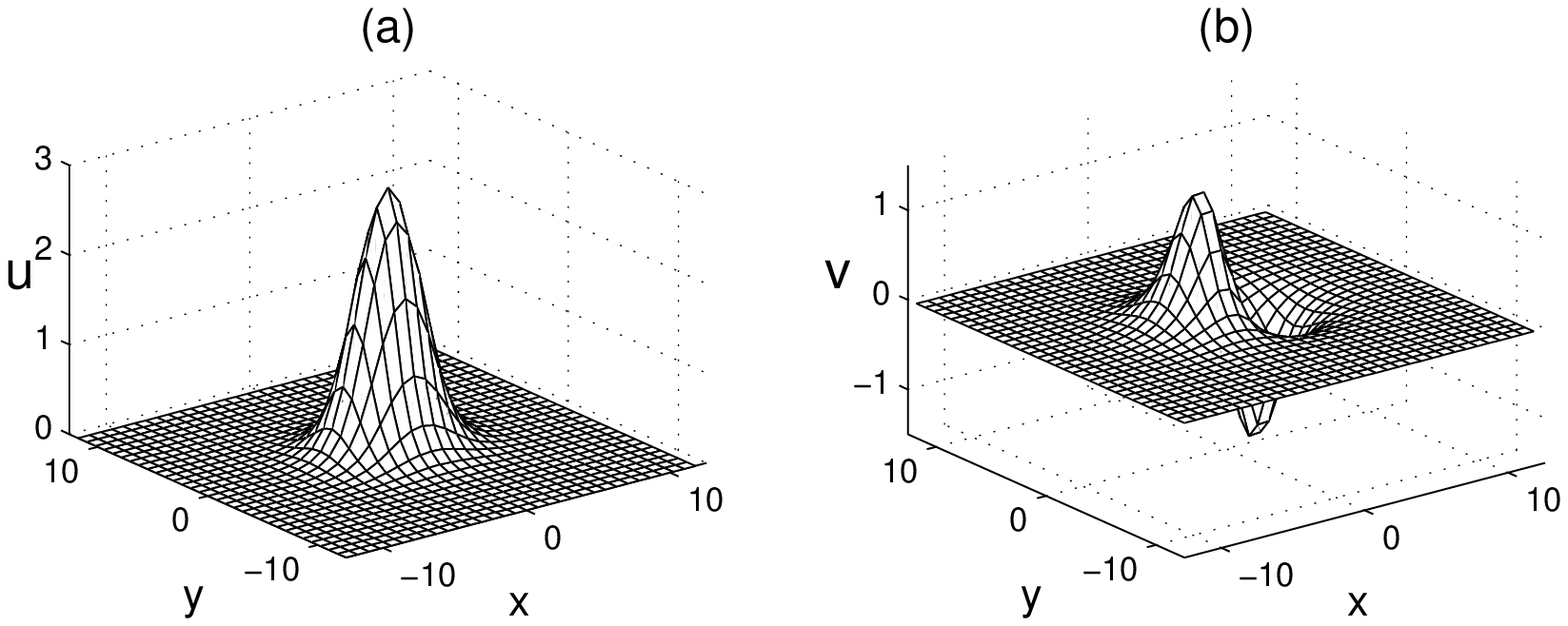}{1.0}}

\parbox{6cm}{\postscript{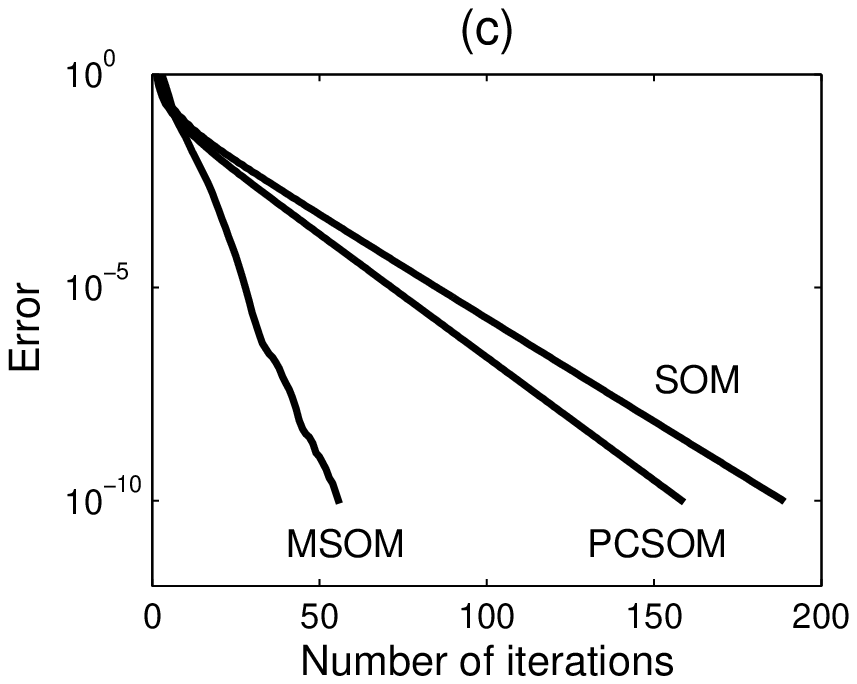}{1.0}}
\caption{(a, b) A dipole-mode vector solitary wave in the coupled NLS system (\ref{CPNLS1})-(\ref{CPNLS2}) at
$\mu_1=1$ and $\mu_2=0.5$ ($P_1=85.3884, P_2=29.1751$);
(c) error diagrams for SOM (\ref{SOM}), MSOM (\ref{MSOM}), (\ref{choice2}) and PCSOM
(\ref{PCSOM_vector})-(\ref{PCSOM3_vector}) at optimal scheme parameters (see text).   }
\end{center}
\end{figure}

\begin{figure}[h]
\begin{center}
\parbox{12cm}{\postscript{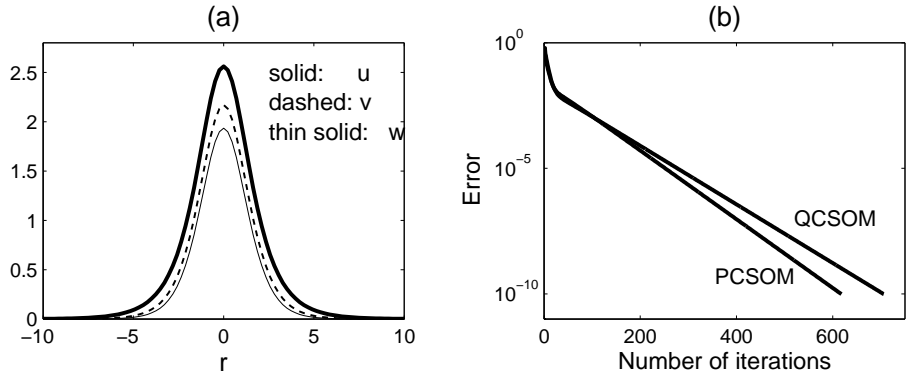}{1.0}}

\caption{(a, b) A fundamental soliton in the three-wave model
(\ref{threewave1})-(\ref{threewave3}) at $\mu_1=0.5$ and $\mu_2=1$,
where $Q_1=\langle u, u\rangle+\langle w, w\rangle=66.3096$, and
$Q_2=\langle v, v\rangle+\langle w, w\rangle=47.2667$; (c) error
diagrams for PCSOM (\ref{g_normstep})--(\ref{muQC}) and QCSOM
(\ref{QCSOM}) at optimal scheme parameters (see text).
\label{threewavefig} }
\end{center}
\end{figure}

\begin{figure}[h]
\begin{center}
\parbox{12cm}{\postscript{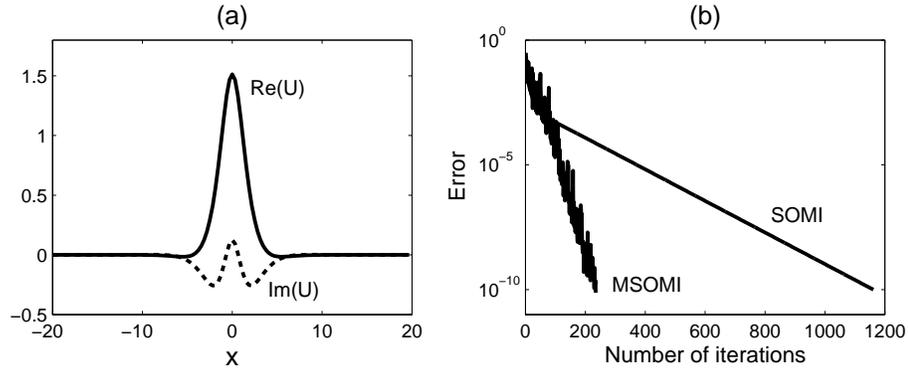}{1.0}}

\caption{ (a) An isolated solitary wave in the Ginzburg-Landau
equation (\ref{GL}) with $\gamma_0=0.3$ and $\gamma_1=1$; (b) error
diagrams of SOMI (\ref{SOMI1})-(\ref{SOMI2}) and MSOMI
(\ref{MSOMI1})-(\ref{choice3}) at optimal $c$ and $\Delta t$ values (see text).
\label{fig_GL}}
\end{center}
\end{figure}

\end{document}